\documentclass[12pt,english]{article}
\usepackage[english]{babel}
\usepackage[dvips]{graphicx}
\usepackage{epsfig} % for postscript graphics files

\usepackage{amsmath} % assumes amsmath package installed
\usepackage{amssymb}  % assumes amsmath package installed
\usepackage{subfig}
\usepackage{psfrag}

\newtheorem{example}{Example}[section]

\newtheorem{theorem}{Theorem}[section]

\newcommand{\SNR}{\mathrm{SNR}}
\newcommand{\SINR}{\mathrm{SINR}}
\newcommand{\BS}{\mathrm{BS}}
\renewcommand{\Omega}{\mathcal{D}}

\title{\LARGE \bf
Spatial games and global optimization for  mobile association problems\footnote{
Part of this work has been presented at the 49th IEEE Conference on Decision and Control.
}
}

\author{Alonso Silva\thanks{Alonso Silva and Eitan Altman are with INRIA. {\tt\small Email: Name.Surname@inria.fr}}~,
Hamidou Tembine\thanks{Hamidou Tembine and M\'erouane Debbah are with Sup\'elec. {\tt\small Email: Name.Surname@supelec.fr}}~,
\\
Eitan Altman$^\dag$, M\'erouane Debbah$^\ddag$% <-this % stops a space
}

\begin{document}
%\pagewiselinenumbers

\maketitle
%\thispagestyle{empty}
%\pagestyle{empty}

%%%%%%%%%%%%%%%%%%%%%%%%%%%%%%%%%%%%%%%%%%%%%%%%%%%%%%%%%%%%%%%%%%%%%%%%%%%%%%%%%%%%%%%%%%%%%%%%%%%%%%

%In this work, we focus on the mobile association problem:
%the determination of the cells corresponding to each base station,
%i.e., the locations at which intelligent mobile terminals prefer
%to connect to a given base station rather than to others.
%This work proposes an approach based on optimal transport theory
%to characterize the solution based  on fluid approximations.
%We characterize the optimal solution from both the global network
%and the individual user points of view.

\begin{abstract}
The basic optimal transportation problem consists in finding the most
effective way of moving masses from one location to another, while
minimizing the transportation cost. Such concept has been found to be
useful to understand various mathematical, economical, and control
theory phenomena, such as Witsenhausen's counterexample in stochastic
control theory, principal-agent problem in microeconomic theory,
location and planning problems, etc.\newline
%It has also applications in engineering design problems, image processing,
%crowd dynamics and collective motion.
\noindent In this work, we focus on mobile association problems: the
determination of the cells corresponding to each base station, i.e.,
the locations at which intelligent mobile terminals prefer to connect
to a given base station rather than to others. This work combines game
theory and optimal transport theory to
characterize the solution based  on fluid approximations. We
characterize the optimal solution from both the global network and the
mobile user points of view.
\end{abstract}
%%%%%%%%%%%%%%%%%%%%%%%%%%%%%%%%%%%%%%%%%%%%%%%%%%%%%%%%%%%%%%%%%%%%%%%%%%%%%%%%%%%%%%%%%%%%%%%%%%%%%%%
\section{Introduction}

Future wireless networks will be composed by intelligent
mobile terminals, capable of accessing multiple radio access technologies,
and able to decide for themselves the wireless access technology
to use and the access point to which to connect.
Within this context, we study the mobile association problem,
where we determine the locations at which intelligent mobile terminals prefer
to connect to a given base station rather than to others.
We consider that these capabilities should be taken into account
in the design and strategic planning of wireless networks.
We analyze the case where mobile terminals within a cell share the same spectrum,
and consequently, mobile terminals' decisions to which base station to connect
affects the decision process of the other mobile terminals in the network.
From these interactions, mobile terminals learn their optimal access point (self-learning),
where the optimality depends upon the context.

Starting from the seminal paper of Hotelling~\cite{hotelling}
a large area of research on location games has been developed.
In~\cite{hotelling}, the author introduced the notion of spatial competition in a duopoly situation.
Plastria~\cite{plastria} presented an overview of the research on locating
one or more new facilities in an environment where
competing facilities already exist.
Gabszewicz and Thisse~\cite{gabszewicz} provided another general survey on location games.
Altman et al.~\cite{spatial} studied the duopoly situation in the uplink scenario of a cellular network
where the users are placed on a line segment. The authors realized that, considering the particular cost structure
that arises in the cellular context, complex cell shapes are obtained at the equilibrium.
Our work focuses on the downlink scenario and in a more general situation where a finite number of base
stations can compete in a one-dimensional and two-dimensional case
without making any assumption on the symmetry of the users location.
In order to do that, we propose a new framework for the mobile association problem using optimal transport theory
(see~\cite{villani} and references therein).
This theory was pioneered by Monge~\cite{monge} and Kantorovich~\cite{kantorovich} and it has been proven to be useful
in many mathematical, economical, and control theory contexts~\cite{buttazzo2005,carlier2004,carlier2007}.
%as well as in the road traffic community~\cite{carlier}.
There is a number of works on ``optimal transport''~(see~\cite{yu},
and references therein)
however the authors in \cite{yu} consider only an optimal selection of routes
but do not use the rich theory of optimal transport.
The works on stochastic geometry are similar
to our analysis of wireless networks (see e.g.~\cite{baccelli} and references therein)
but in our case we do not consider any particular deployment distribution function.
Fluid models allow us to have this general deployment distribution function.

%% Contributions %%%%%%%%%%%%%%%%%%%%%%%%%%%%%%%%%%%%%%%%%%%%%%%%%%%%%%%%%%%%%%%%%%%%%%%%%%%%%
In the current work,
we determine the spatial locations at which intelligent
mobile terminals would prefer to connect to a given base station rather than to other
base stations in the network.
We obtain as well the spatial locations
which are more convenient from a global and centralized point of view.
Obviously, in both approaches the optimality depends upon the context.
In both considered cases, our aim is
the minimization of the total power
of the network,
which can be considered as an energy-efficient objective,
while maintaining a certain level of throughput 
for each user connected to the network.
We propose  the rich theory of optimal transport as the main tool of modelization of these mobile association problems.
We are able to characterize these mobile associations under different policies
and give illustrative examples of this technique.
%%%%%%%%%%%%%%%%%%%%%%%%%%%%%%%%%%%%%%%%%%%%%%%%%%%%%%%%%%%%%%%%%%%%%%%%%%%%%%%%%%%%%%%%%%%%%%%%%%%%%

The remaining of this paper is organized as follows.
Section~\ref{model} outlines the problem formulation of minimizing the total network power under quality of service constraints.
% from different perspectives.
% In Section~\ref{secbasics} we give some basics in optimal transport theory.
We address the problem %of minimizing the power under quality of service constraint
for the downlink case. Two different policies are studied:
round robin scheduling policy (also known as time fair allocation policy) and rate fair allocation policy
defined in Section~\ref{model} and studied in detail in Section~\ref{sectionroundrobin} and Section~\ref{sectionratefair}
with uniform and non-homogeneous distribution of users.
In Section~\ref{secnum} we give numerical examples for one-dimensional and two-dimensional mobile terminals deployment distribution functions.
Section~\ref{secconclusion} concludes the paper.

\renewcommand{\lambda}{f}
\section{The system model and problem formulation}\label{model}

%\begin{multicols}{2}

\begin{table}
\renewcommand{\arraystretch}{1.8}
\caption{Notation}
\label{notation}
\centering
\begin{tabular}{|c|l|}
    \hline
    $N$ & Total number of MTs in the network\\ \hline
    $K$ & Total number of BSs\\ \hline
    $\lambda$ & Deployment distribution of MTs\\ \hline
    $(x_i,y_i)$ & Position of the $i$-th BS\\ \hline
    $C_i$ & Cell determined by the $i$-th BS\\ \hline
%    $\mathrm{BS}_i$ & Base station that determines its cell $C_i$\\ \hline
    $N_i$ & Number of MTs associated to the $i$-th BS\\ \hline
    $M_i$ & Number of carriers offered by the $i$-th BS\\ \hline
%  \end{tabular}
% \end{center}
%\begin{center}
%
%  \begin{tabular}{|l|l|}
%    \hline
%    $D$ & Geographical reference of the network\\ \hline
    $\kappa_i$ & Penalization function of non-service\\ \hline
    $h_i$ & Channel gain function in the $i$-th cell\\ \hline
    $\xi_i$ & Path loss exponent in the $i$-th cell\\ \hline
  \end{tabular}
\end{table}
%\end{multicols}

A summary of the notation used on this work can be found in Table~\ref{notation}.
We consider a %bounded domain~$D$ of the plane which is the geographic reference for the network
network deployed on a region, denoted by~$\Omega$, over the two-dimensional plane.
%which will be the geographical reference of the network.
The mobile terminals (MTs) are distributed %.
%and consider that the network contains
%large number of mobile terminals % square integrable %\footnote{The exact condition is square integrable in the domain so
%the examples~()(\ref{Doumbodo}) is included in our work.}
%distributed
%or with a finite number of discontinuities
according to a given deployment distribution function~\mbox{$\lambda(x,y)$}.
To fix ideas, if the considered region is a square \mbox{$\Omega=[0~\textrm{km},1~\textrm{km}]\times[0~\textrm{km},1~\textrm{km}]$}
and the distribution of the users is uniform~\mbox{$\lambda(x,y)=1$}, then
the proportion of users in the sub-region \mbox{$\mathcal{A}=[0~\textrm{km},1/2~\textrm{km}]\times[0~\textrm{km},1/2~\textrm{km}]$}
is
\begin{equation}
\frac{
\int\!\!\!\int_\mathcal{A} \lambda(x,y)\,dx\,dy
}{
\int\!\!\!\int_\mathcal{D} \lambda(x,y)\,dx\,dy
}=
\frac{
\mathrm{Area}(\mathcal{A})
}{
\mathrm{Area}(\mathcal{D})
}=
\frac{1}{4}.
\end{equation}

The first equality is obtained because the distribution of the users is uniform.
However, the expression at the left-hand side is general and it is always equal to
the proportion of mobiles in a sub-region~$\mathcal{A}$.
To simplify the notation, we normalize the function~$\lambda$ such that
\begin{equation}\label{eq:normalization}
\int\!\!\!\!\int_\mathcal{D} \lambda(x,y)\,dx\,dy=1.
\end{equation}
Consequently, the function~$\lambda$ is a measure of the proportion of users over the network.

%%%%%%%%%%%%%%%%%%%%%%%%%%%%%%%%%%%%%%%%%%%%%%%%%%%%%%%%%%%%%%%%%%%%%%%%%%%%%%%%%%%%%%%%%%%%%%%%
The number of MTs in a subset of the network area~$A\subseteq\Omega$, denoted by~$N(A)$,
is given by
\begin{equation}
N(A)=N\left(\int\!\!\!\int_A\lambda(x,y)\,dx\,dy\right),
\end{equation}
where~$N$ is the total number of MTs in the network.
The integral on the right hand side between brackets takes into account the
proportion of MTs distributed over the network area~$A$.
%%%%%%%%%%%%%%%%%%%%%%%%%%%%%%%%%%%%%%%%%%%%%%%%%%%%%%%%%%%%%%%%%%%%%%%%%%%%%%%%%%%%%%%%%%%%%%%%%

Examples of distribution of the users~$\lambda(x,y)$:
\begin{enumerate}
\item\label{uniform} If the users are distributed uniformly over the network,
then the measure of the proportion of the users is given by
\begin{equation}\label{eq:constant}
\lambda(x,y)=\frac{K}{\lvert\Omega\rvert},
\end{equation}
where $\lvert\Omega\rvert$ is the total area of the network, and $K$ is a coefficient of normalization so
that equation~\eqref{eq:normalization} holds. In this particular case, $K\equiv 1$.
\item\label{Doumbodo} If the users are distributed according to different levels of population density,
then the measure of the proportion of the users would be
\begin{equation}
\lambda(x,y)=
\left\{
\begin{array}{l}
\lambda_{\rm HD}\text{ if $(x,y)$ is at a High Density region},\\
\lambda_{\rm ND}\text{ if $(x,y)$ is at a Normal Density region},\\
\lambda_{\rm LD}\text{ if $(x,y)$ is at a Low Density region},
\end{array}
\right.
\end{equation}
\noindent where~$\lambda_{\rm HD} > \lambda_{\rm ND} > \lambda_{\rm LD}$
are defined similarly to equation~\eqref{eq:constant} with constants of normalization~$K_{\rm HD}$, $K_{\rm ND}$, $K_{\rm LD}$,
such that $K_{\rm HD} > K_{\rm ND} > K_{\rm LD}$. As a particular example, we could consider
in the domain $[0,1]\times[0,1]$
as High Density region the area $[0,1/4)\times[0,1]$ with $K_{\rm HD}=2$,
as Normal Density region the area $[1/4,1/2)\times[0,1]$ with $K_{\rm ND}=1$,
and as Low Density region the area $[1/2,1]\times[0,1]$ with $K_{\rm LD}=1/2$.
In that case, equation~\eqref{eq:normalization} holds.

\item\label{radial} If the distribution of the users is radial with more mobile terminals in the center
of the network area and less mobile terminals in the suburban areas then
\begin{equation}
\lambda(x,y)=\frac{R_D^2-(x^2+y^2)}{K_D},
\end{equation}
where $R_D$ is the radius of the network and~$K_D$
is a coefficient of normalization.
%\item If the distribution of users is a Poisson process with intensity~$\nu$,
%then
%\[
%\lambda(x,y)=e^{-\nu\pi r^2}
%\]
%where $r$ is the polar coordinate representation of~$(x,y)$.
%This particular case has been examinated in~\cite{baccelli}.
\end{enumerate}
Notice that the distribution of users~$\lambda(x,y)$ considered in our work
is more general than all the examples mentioned above.

In the network, we consider $K$~base stations (BSs), denoted by \mbox{$\BS_i\,,i\in\{1,\ldots,K\}$}, % $\BS_ 1,\BS_2,\ldots,\BS_K$
located at the fixed positions \mbox{$(x_i,y_i)\,,i\in\{1,\ldots,K\}$}. % $(x_1,y_1),(x_2,y_2)\ldots,(x_K,y_K)$.
The interference between the different BS signals is ignored.
We assume that the neighbouring BSs transmit their signals in
orthogonal frequency bands.
Furthermore, we assume that
interference between BSs that are
far from each other is negligible.
Consequently, instead of considering the $\SINR$ (Signal to Interference plus Noise Ratio), we consider as performance measure the~$\SNR$ (Signal to Noise Ratio).
%% % % % %Our objective is to determine the optimal mobile association to each base station
%% % % % %in order to minimize the total power of the network needed to maintain a quality of service,
%% % % % %in this case, an average throughput of~$\bar\theta(x,y)>0$
%% % % % %for each mobile of the network located at position~$(x,y)$.
%% % % % %We also analyse and determine the equilibrium situation where each of the mobile terminals
%% % % % %decide in itself with which base station to connect in order to maximize its own rate. %throughput.
%% % % % %
%% % % % %
%
%% The generalized function may take into account the throughput or the output or the
%
%% \subsection{Downlink case}
%
We consider the downlink case (transmission from base stations to mobile terminals) and assume
that each BS is going to transmit only to MTs associated to it.
We denote by~$C_i$ the set of mobiles associated to the $i$-th BS,
and by~$N_i$ the number of mobiles within that cell, both quantities to be determined.
Notice that since the distribution of users~$\lambda(x,y)$ considered in our work
is general, instead of considering a particular distribution of mobiles, that we denote~$\tilde\lambda(x,y)$, and an average throughput, 
that we denote~$\bar\theta(x,y)$,
in each location~$(x,y)$,
we can consider a constant average throughput~$\theta>0$ and we can vary the distribution of mobiles~$\lambda(x,y)$ such that the following equation holds:
\begin{equation}
\lambda(x,y)\theta=\tilde\lambda(x,y)\bar\theta(x,y).
\end{equation}
This would simply translate in the fact that for example mobile terminals with double demand than others
would be considered as two users with the same demand.
This can be done because of the fluid approximation of the network.

If the number of mobiles  is greater than the maximum number of carriers available in the $i$-th cell, denoted by~$\mathrm{MAX}_i$,
we consider a penalization cost function given by
\begin{equation}
\kappa_i(N_i)=
\left\{
\begin{array}{cl}
0&\textrm{ if }N_i\leq\mathrm{MAX}_i,\\
\bar\kappa_i(N_i-\mathrm{MAX}_i)&\textrm{ if }N_i> \mathrm{MAX}_i.
\end{array}
\right.
\end{equation}
We assume that $\bar\kappa_i$~can be either a constant or a non-decreasing
function\footnote{For example, the
maximum number of possible carriers in WiMAX
is around~$2048$, so by using this technology we have~$\mathrm{MAX}_i=2048$.}.
We first study the case~$N_i\leq M$
and we study the general case in Section~\ref{subsec:penalization}.

% {\bf Discussion about the pertinency of taking into account this assumption ?}
%hen the total number of mobiles in the network, denoted by~$N$, will be given by~$N(D)=\sum_{i=1}^K N_i(C_i)$.
The power transmitted from~$\BS_i$ to an MT located at position~$(x,y)$,
is denoted by~$P_i(x,y)$. The received power at an MT served by~$\BS_i$ is~$P_i(x,y) h_i(x,y)$.
%\footnote{Vectors are denoted by bold fonts}
%from base station $\BS_i$ is given by $P_i(x,y) h_i(x,y)$
%where $h_i(x,y)$ is the channel gain.
We shall further assume that the channel gain corresponds to the path loss given by
\mbox{$ %\begin{equation}\label{gain}
h_i(x,y)=(\sqrt{R^2+d^2_i(x,y)})^{-\xi}
$} %\end{equation}
where $\xi$ is the path loss exponent, $R$ is the height of the base station, and $d_i(x,y)$ is the distance
between a MT at position~$(x,y)$ and~$\BS_i$ located at~$(x_i,y_i)$, {\it i.e.},
\mbox{$d_i(x,y)=\sqrt{(x_i-x)^2+(y_i-y)^2}$}.
%We assume for the downlink case that
%between neighboring base station they transmit in
%orthogonal channels
%and the interference between those cells that are
%far from each other
%is negligible,
%so instead of considering the $\SINR$ (Signal to Interference plus Noise Ratio) we only consider the~$\SNR$ (Signal to Noise Ratio). For the uplink case we
%will consider the interference between mobile terminals so we will consider the~$\SINR$ (Signal to Interference plus Noise Ratio).
The~$\SNR$ %(Signal to Noise Ratio)
received at mobile terminals at position~$(x,y)$ in cell~$C_i$ is given by
$ %\begin{equation}\label{SNR}
\SNR_i(x,y)=P_i(x,y) h_i(x,y)/\sigma^2,
$ % \end{equation}
where~$\sigma^2$ is the noise power.
We assume that the instantaneous mobile throughput % of mobiles
is given by the following expression, which is based
on Shannon's capacity theorem:
\begin{equation}
\theta_i(x,y)=\log(1+\SNR_i(x,y)).
\end{equation}

We want to satisfy an average throughput for
MTs located at position~$(x,y)$ given by~\mbox{$\bar\theta(x,y)>0$}.
We shall consider for this objective two policies defined in~\cite{kasbekar}:
\begin{itemize}
\item[(A)]~{\sl Round robin scheduling policy:} where each~$\BS$
devotes an equal fraction of time for the transmission to each MT
associated to it, and
\item[(B)]~{\sl Rate fair allocation policy:} where each base station~$\BS$ maintains a constant power %~$P$
sent to the mobile terminals within its cell and modifies the fraction of time
allowed to mobile terminals with different channel gains, such that the average
transmission rate demand is satisfied.
\end{itemize}

For more information about this type of policies in the one dimensional case,
see~\cite{kasbekar}.

\subsection{Round robin scheduling policy: Global Optimization}

Following this policy,~$\BS_i$
devotes an equal fraction of time for transmission to MTs
located within its cell. % From equation~\eqref{mobilenumber}
The number of MTs located in the $i$-th cell is~\mbox{$N_i$}, to be determined together with the cell boundaries.
% As the constraint~\eqref{qos} is reached, then the throghput of each mobile located in~$(x,y)$
As $\BS_i$ divides its time of service proportional to the quantity of users %~$N_i(C_i)$
within its cell, then the throughput is given by
%\[
%\overline{\theta}_i(x_0,y_0)=\frac{1}{N_i}\log(1+\overline{\SNR}_i(x_0,y_0)).
%\]
%%%%% replaced by
\mbox{$
\theta_i^{\mathrm{RR}}(x,y)=\theta_i(x,y)/N_i. %\frac{1}{N_i}\log(1+\SNR_i(x,y)).
$}
In order to satisfy a throughput $\bar\theta(x,y)$, \mbox{$\theta^{\mathrm{RR}}(x,y)\geq\bar\theta(x,y)$},
or equivalently, in terms of the power %the~$\SNR$~$\theta(x,y)$ will be
\begin{equation}
P_i(x,y)\geq\frac{\sigma^2}{h_i(x,y)}(2^{N_i\bar\theta(x,y)}-1).
\end{equation}
As our objective function is to minimize the total power of the network,
the constraint will be reached, and we obtain by replacement
\begin{equation}\label{qualityofservice}
%\mbox{$ %\begin{equation}\label{pointpower}
% P_i(x,y)=\sigma^2 N_i\theta(x,y)(R^2+d_i^2(x,y))^{\xi/2}
P_i(x,y)=\sigma^2 (2^{N_i\bar\theta(x,y)}-1)([R^2+d_i^2(x,y)]^{1/2})^{+\xi}.
\end{equation}%$} %\end{equation}

From last equation we observe that:
a) if the quantity of mobile terminals increases within the cell, %~$C_i$,
% as the base station~$C_i$ is dividing its time slots with respect to the number
% of the mobiles within the cell in order to satisfy the average throughput~$\theta$ of the mobile terminals
the base station will need to transmit more power to each of the mobile terminals.
The reason to do that is because the base station %~$\BS_i$
is dividing each time-slot into mini-slots with respect to the number
of the mobiles within its cell, and %~$C_i$.%, in order to satisfy the average throughput~$\theta(x,y)$ of the mobile terminals.
%\item The function~$N_i$ give us the dependence of the power cost function % of the power
%with respect to the quantity of mobile terminals~$N_i$ within the cell~$C_i$. If there are more nodes
%inside the cell~$C_i$, as the base station~$C_i$ is dividing its time slots with respect to the number
%of the mobiles within the cell in order to satisfy the average throughput~$\theta$ of the mobile terminals
%it will need to transmit more power to each of the mobile terminals.
%to satisfy the average throughput~$\theta$ of the mobiles.
b) the function~$(R^2+d_i^2(x,y))^{\xi/2}$ on the right hand side give us the dependence
of the power with respect to the distance between the base station %~$\BS_i$
and the mobile terminal located at position~$(x,y)$.% inside the cell. %located at~$(x,y)$ inside the cell~$C_i$.
% \end{itemize}

% The problem that we are trying to solve deals with
Our objective is to find the optimal mobile association in order to minimize the total
power of the network.
Then as the total power
\begin{gather}
P_{\rm total}=\sum_{i=1}^K P_i^{\rm intra},\\
\textrm{where}\quad P_i^{\rm intra}=\int\!\!\!\!\int_{C_i}P_i(x,y)\lambda(x,y)\,dx\,dy,
\end{gather}
$P_i^{\rm intra}$ is the intracell power consumption in cell $C_i$.
% from previous equation the problem reads

The global optimization for the mobile association problem is to determine the cells~\mbox{$C_i,\,i\in\{1,\ldots,K\}$},
to minimize the total power of the network:
\begin{equation}\label{roundrobin}
(\mathrm{RR})\quad\min \sum_{i=1}^K\int\!\!\!\!\int_{C_i}P_i(x,y)\lambda(x,y)\,dx\,dy,
\end{equation}
subject to~\eqref{qualityofservice}, where~$\lambda(x,y)$ is the deployment distribution function of the users.
We solve this problem in Section~\ref{sectionroundrobin}.

\subsection{Generalized $\alpha$-fairness formulation for minimization problems}

The general formulation for the problem of maximization of a function of the throughput given the
constraint on the maximal power used admits a generalized $\alpha$-fairness formulation given by:
\begin{equation}
\max\ \sum_j \frac{1}{1-\alpha}[f(\theta_j)^{1-\alpha}-1],
\end{equation}
where we can identify different problems for different values of $\alpha$:
\begin{itemize}
\item $\alpha=0$ maximization of throughput problem
\item $\alpha\rightarrow 1$ proportional fairness (a uniform case of Nash bargaining)
\item $\alpha=2$ delay minimization
\item $\alpha\rightarrow +\infty$ maxmin fairness (maximize the minimum throughput that a user can have).

\end{itemize}

Since in minimization problems, the formulation is different since we are minimizing the total power on the network
given the constraint of a minimum level of throughput, we define the following formulation, that we
call generalized $\alpha$-fairness for minimization problems:
\begin{equation}
\min\ \sum_i \int_{C_i}\frac{1}{\alpha-1}[f(P_i(x,y))^{\alpha-1}-1]\lambda(x,y)\ dx \ dy,
\end{equation}
where we can also identify different problems for different values of $\alpha$:
\begin{itemize}
\item $\alpha=0$ maximization  of the inverse of power (energy efficiency maximization)
\item $\alpha\rightarrow 1$ proportional fairness
\item $\alpha=2$ minimization of total power
\item $\alpha\rightarrow +\infty$ minmax fairness (to minimize  the maximum power per BS).
\end{itemize}

Note: the minmax fairness is not well studied in the literature but one can map
the maxmin fairness studies into the minmax fairness for minimization problem.
The convexity properties required becomes concavity, Schur convexity, sub-stochastic ordering, etc.

%%%%%%%%%%%%%%%%%%%%%%%%%%%%%%%%%%%%%%%%%%%%%%%%%%%%%%%%%%%%%%%%%%%%%%%%%%%%%%%%%%%%%%%
\subsection{Rate fair allocation policy: User Optimization}
% \begin{itemize}
% \item User optimization
% \end{itemize}
% In the round robin scheduling policy each base station~$\BS_i$ modifies
% the power sent to mobile terminals with different channel gains
% in order to satisfy a throughput of~$\theta(x,y)$ for each mobile
% located at position~$(x,y)$.
In the rate fair allocation policy, each~$\BS$ will
maintain a constant power sent to MTs within its cell,
{\it i.e.},
$ %\begin{equation}\label{power}
P_i(x,y)=P_i$ for each MT at location $(x,y)$ inside cell $C_i$.
% \end{equation}
However, the $\BS$ modifies the fraction of time allotted to
MTs, set in such a way that the average transmission rate to
each MT with different channel gain is the same, denoted by~$\bar\theta(x,y)$, for each mobile
located at position~$(x,y)$.

%  Let~$r_i$ be the fixed rate of MTs located inside cell~$C_i$.
%  Following the rate fair allocation policy, the rate~$r_i$
%  is given by (see e.g.~\cite{kasbekar}):
%  the fraction of time
%  that MT at position~$(x,y)\in C_i$ receives positive throughput is
% \[
% \frac{r_i}{\log(1+\SNR_i(x))}.
% \]
%   %%% replaced by
%   $
%   \frac{r_i}{\SNR_i(x,y)}.
%   $
%   Then, the fixed rate~$r_i$ is the solution to equation
%   %\[
%   %\int\!\!\!\!\int_{C_i}\frac{r_i}{\log(1+\SNR_i(x,y))}\lambda(x,y)\,dx\,dy=\theta.
%   %\]
%   %%% replaced by
%   $
%   \int\!\!\!\int_{C_i}\frac{r_i}{\SNR_i(x_0,y_0)}\lambda(x,y)\,dx\,dy=\beta,
%   $
%  where~$\beta$ is the fraction of time  guaranteed from the regulator to have no interference from the other BSs. Then, the rate
%  %\[
%  %r_i=\theta\Big(\int\!\!\!\!\int_{C_i}\frac{1}{\log(1+\SNR_i(x,y))}\lambda(x,y)\,dx\,dy\Big)^{-1}
%  %\]
%  %%% replaced by
%  \begin{equation}
%  r_i=\Big(\int\!\!\!\!\int_{C_i}\frac{1}{\log(1+\SNR_i(x,y))}\lambda(x,y)\,dx\,dy\Big)^{-1}\beta.
%  \end{equation}
%
%From equations~\eqref{SNR} and~\eqref{power} replacing the~$\SNR$ we obtain
%\[
%r_i=\Big(\int\!\!\!\!\int_{C_i}\frac{\sigma^2}{P_ih_i(x,y)}\lambda(x,y)\,dx\,dy\Big)^{-1}\beta,
%\]
%and from equation~\eqref{gain} we obtain
% Replacing expressions from previous calculations we obtain
% \begin{equation}\label{rate}
% r_i=\beta P_i\Big(\int\!\!\!\!\int_{C_i}\sigma^2(R^2+d_i(x,y))^{\xi/2}\lambda(x,y)\,dx\,dy\Big)^{-1},
% \end{equation}

We study the equilibrium states where each MT chooses the BS
which will serve it.
Given the interactions with the other mobile terminals it
doesn't have any incentive to unilaterally change its strategy.
A similar notion of equilibrium has been studied
in the context of large number of small players
in road-traffic theory by Wardrop~\cite{Wardrop}.
%As we are working with a large number of small players similar type of equilibrium has been studied
%A Wardrop equilibrium is the analog of a Nash equilibrium
%for the case of a large number of small players,
%as in the case with mobiles in our setup.

{\bf Definition.-} The Wardrop equilibrium is given in the context of cellular systems by:
\begin{subequations}
\begin{equation}\label{wardropalice}
\textrm{If }\int\!\!\!\!\int_{C_i}\lambda(x,y)\,dx\,dy>0,
\textrm{ then } \theta_i=\max_{1\leq j\leq K} \theta_j(C_j),%\quad\textrm{for all}\quad 1\leq i\leq K,
\end{equation}
\begin{equation}\label{wardropbob}
\textrm{else if }\int\!\!\!\!\int_{C_i}\lambda(x,y)\,dx\,dy=0,
\textrm{ then }\theta_i\leq\max_{1\leq j\leq K} \theta_j(C_j).
\end{equation}
\end{subequations}

A Wardrop equilibrium is the analog of a Nash equilibrium
in the case of a large number of small players,
where, in our case, we consider the mobile terminals as the players.
In this setting, the Wardrop equilibrium
indicates that if there is a positive proportion of mobile terminals associated to the $i$-th base station
(the left-hand side condition in~\eqref{wardropalice}),
then the throughput that the mobile terminals obtain
is the maximum that they would obtain from any other base station (right-hand side consequence in~\eqref{wardropalice}).
The second condition indicates that if there is one base station that doesn't have any
mobile terminal associated to it (left-hand side condition in~\eqref{wardropalice}),
it is because the mobile terminals can obtain a higher throughput by connecting to one of the other base stations
(right-hand side consequence in~\eqref{wardropalice}).

We assume that each base station is serving at least one mobile terminal,
(if that is not the case, we remove the base station that is not serving any mobile
terminal). Then, the equilibrium situation is given by
\begin{equation}\label{eq:condition}
\theta_1=\theta_2=\ldots=\theta_K.
\end{equation}
To understand this equilibrium situation, consider as an example the simple case of two base stations:~$\BS_1$ and~$\BS_2$.
Assume that~$\BS_1$ offers more throughput than~$\BS_2$.
Then, the mobile terminals being served by~$\BS_2$ will have an incentive to
connect to~$\BS_1$. %Notice that the terms inside the integral
%of equation~\eqref{rate} are all positive.
%Then the rate transmitted from
The transmitted throughput depends inversely on the quantity of mobiles connected to the base station.
%since it depends on the quantity of mobiles through the size of the cell~$C_i$ and through
%the density of mobiles inside the cell~$\lambda(x,y)$.
As more mobile terminals
try to connect to base station~$\BS_1$ the throughput will diminish until
arrive to the equilibrium where both base stations will offer the same throughput.

The condition $\theta_1=\theta_2=\ldots=\theta_K$ is equivalent in our setting to
the condition $r_1=r_2=\ldots=r_K$,
and also to the condition $\SNR_1=\SNR_2=\ldots=\SNR_K$.
Let us denote by $r$ to the rate offered by the base station at equilibrium, {\sl i.e.},
$r:=r_1=r_2=\ldots=r_K$ and
%
%Then, as long as we are in the low-$\SNR$ regime,
%by using the low-$\SNR$ approximation
%of throughput, i.e.,
%the throughput of a mobile at~$\mathbf{x}$
%is $\log(1+\SNR)\approx\SNR$, we have that:
%from equation~\eqref{rate}
%
we denote by $\beta$ to the $\SNR$ offered by the base station at equilibrium,  {\sl i.e.},
$\beta=\SNR_1=\SNR_2=\ldots=\SNR_K$.

This condition in terms of the power is equivalent to
\begin{equation}\label{powercell}
P_i^{\rm intra}=\beta\int\!\!\!\!\int_{C_i}\sigma^2([R^2+d^2_i(x,y)]^{1/2})^{+\xi}\lambda(x,y)\,dx\,dy.
\end{equation}

We want to choose the optimal mobile assignment in order to minimize the total power of the network under the constraint
that the mobile terminals have an average throughput of~$\theta$, {\it i.e.},
\begin{equation}
\operatorname*{Min}_{C_i} P_{\rm total}=\sum_{i=1}^K P_i^{\rm intra}.
\end{equation}

%Notice that in each cell we can differentiate two regions
%\begin{enumerate}
%\item In one of them the noise~$R$ compared to the distance is very small~{\it i.e.} $R^2<\!\!\!<d^2$.
%\item In the other both have the same order of magnitude. We call this region~$C_i^R$
%\end{enumerate}
%The second case is important for the new field of research called small cells or ecocells.
%For more about this see the references ().

Then our problem reads
\begin{equation}
(\mathrm{RF})\quad\operatorname*{Min}_{C_i}\sum_{i=1}^K\int\!\!\!\!\int_{C_i}\sigma^2(R^2+d^2_i(x,y))^{\xi/2}\lambda(x,y)\,dx\,dy.
\end{equation}

We will solve this problem in Section~\ref{sectionratefair}.
Thanks to optimal transport theory we are able to characterize the partitions considering a
general setting. In the following section, we will briefly describe optimal transport theory
and motivate the solution of the previously considered mobile association problems.

 \section{Basics in Optimal Transport Theory} \label{secbasics}

%%%%%%%%%%%%%%%%%%%%%%%%%%%%%%%%%%%%%%%%%%%%%%%%%%%%%%%%%%%%%%%%%%%%%%%

The theory of mass transportation, also called optimal transport theory, goes back to the original works by Monge
in 1781~\cite{monge}, and later in 1942 by Kantorovich~\cite{kantorovich}.
The work of Brenier~\cite{brenier}
has renewed the interest for the subject and since then
many works have been done in this topic~(see~\cite{villani} and references therein).

The original problem of Monge can be interpreted as the question:
\begin{quote}
``How do you best move given piles of sand
to fill up given holes of the same total
volume?''.
\end{quote}
% Following the notation given by Carlier et al.~\cite{carlier}:
%  Commented for the conference version
In our setting, this problem is of main importance.
Suppose that base stations are sending information to
mobile terminals in a grid area network and the positions of
base stations and mobile terminals are given.
\begin{quote}
What is the ``best move" of information from the MTs to the BSs?
\end{quote}
Both questions share similarities as we will see.
The general mathematical framework to deal with this problem
is a little technical but we encourage to
focus on the main ideas.

The framework is the following:

We first consider a grid area network~$\Omega$ in the one-dimensional case.
As an example,
the function $f(t)$ will represent the
proportion of information sent by mobile terminals
\begin{equation}
d\mu(t):=f(t)\,dt.
\end{equation}
The function $g(s)$ will represent the
proportion of
information
received by a base station
at location~$s$
\begin{equation}
d\nu(s):=g(s)\,ds.
\end{equation}
%\[
%\rho^-(x):=\int_{-\infty}^x g(s)\,ds.
%\]
%  The function $f(t)$ will represent the measure
%  proportion of how much sand can be piled throughput sent by each base station
%  at location~$t$ will send
%  and we denote
%  define
%  Modified by AS
%  \[
%  d\mu(t):=f(t)\,dt.
%  \]
% \[
% d\rho^+(t):=\int_{-\infty}^x f(t)\,dt.
% \]
% will represent the measure of the
% throughput that each base station will send.
The function
$T$ (called transport map) is the function that
transfers information from location~$s$ to location~$t$.
It associates mobile terminals to base stations and transports
information from base stations to mobile terminals.
Then the conditions
that each mobile terminal satisfies its downlink demand is written
\begin{equation}
\int_A F(y)g(y)\,dy=\int_{\{x\in X\,:\,T(x)\in A\}} F(T(x))f(x)\,dx,
\end{equation}
for all continuous function~$F$, where~$X$ is the support\footnote{The support of a function~$f$ is the closure
of the set of points where the function is not zero, {\it i.e.}, support $(f)=\overline{\{t\,:\,f(t)\neq 0\}}$} of function~$f$
and we denote this condition (following the optimal transport theory notation) as
\begin{equation}
T\#\mu=\nu,
\end{equation}
% \[
% T\#\rho^+=\rho^-
% \]
% \[
% f\#\mu^+=\mu^-
% \]
% where $f\#\mu^+$ is defined as
% \[
% \int_0^x f(s)\,ds=\int_0^x g(T(s))\,ds
% \]
% $T\#\rho^+$ is defined as
% \[
% T\#\rho^+(B):=\rho^+(T^{-1}(B))\quad\text{for all}\quad B\subseteq\Omega
% \]
% \[
% f\#\mu^+(B):=\mu^+(f^{-1}(B))\quad\forall B\subseteq X
% \]
which is an equation of conservation of the information.
Notice that, in communication systems there exists
packet loss so in general this constraint may not be satisfied,
but considering an estimation of the packet loss by sending
standard packets test, this constraint can be modified
in the reception measure~$\nu$. If we can not obtain
a good estimation of this reception measure, we can consider
it in its current form as a conservative policy.

In the original problem, %was to move piles of sand to holes,
Monge considered that the cost of moving a commodity from position~$x$ to a position~$y$ depends on the distance
 $c(\lvert x- y\rvert)$. Then the cost of moving a commodity from position~$x$ through~$T$ to its new position~$T(x)$ will be
 $c(\lvert x- T(x)\rvert)$. For the global optimization problem, we consider the additive total cost over the network, which
in the continuum setting will be given by
 \begin{gather}
 \operatorname*{Min}\int_\Omega c(\lvert x-T(x)\rvert)\,f(x)\,dx\ \
 \mathrm{such\; that}\quad T\#\mu=\nu,
% \mathrm{such\; that}\quad T\#\rho^+=\rho^-
 \end{gather}
%
%
%  \begin{gather*}
%  \inf\int_\Omega \lvert x-T(x)\rvert\,d\mu^+(x)\\
%  \mathrm{such\; that}\quad T\#\rho^+=\rho^-
%  \end{gather*}
%
%  \begin{gather*}
%  \inf\int\!\!\!\!\int_D \lvert x-f(x)\rvert\,d\mu^+(x)\\
%  \mathrm{such\; that}\quad f\#\mu^+=\mu^-
%  \end{gather*}
where $\mu$ and $\nu$ are probability measures and
$T:\Omega\to\Omega$ is an integrable function. This
problem is known as Monge's problem in optimal transport theory.%, such that

\begin{figure}[htb]
\centering
\psfrag{MT1}{$MT_1$}
\psfrag{MT2}{$MT_2$}
\psfrag{BS}{$BS$}
\includegraphics[width=10cm,height=6cm]{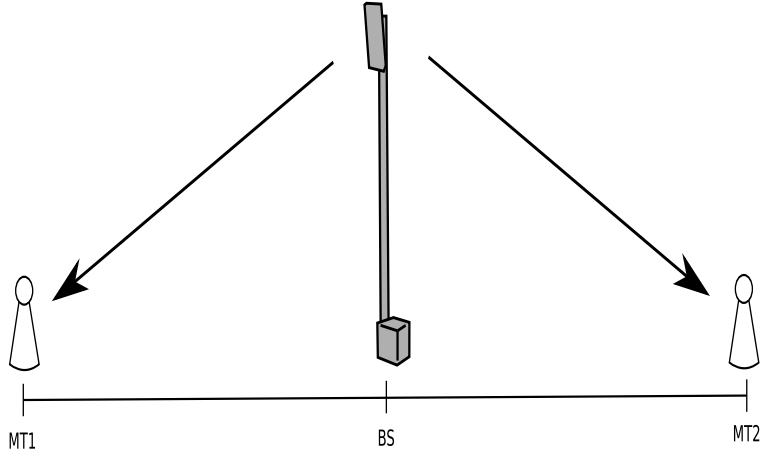}
\caption{Monge's problem can not model a simple scenario of two mobile terminals and one base station.
Kantorovich's problem however can model very general scenarios.}\label{tembine}
\end{figure}

The main difficulty in solving Monge's problem is the
highly non-linear structure of the objective function.
For examples on the limitations on Monge's modelization, see~\cite{villani}.
As an example, consider the domain $\Omega=[0,2]$,
the transmission from a base station located at position~$1$,
denoted $\mu=\delta_1$,
and the throughput demanded to this base station by two mobile terminals 
located at positions~$0$ and~$2$, denoted
$\nu=\frac{1}{2}\delta_{0}+\frac{1}{2}\delta_2$.
%$\mu^-=\frac{1}{2}\delta_{-1}+\frac{1}{2}\delta_1$.
According to the formulation given by Monge,
there is no splitting of throughput, since
everything that is transmitted from one location
has to go to another location.
So this simple problem
doesn't have a transport map
(see~Fig.~\ref{tembine}).
This limitation is due in part to the original considered problem,
but as we will see this limitation is overcome by Kantorovich's approach.
% Given this formulation
We have also pointed out the limitations of Monge's problem
that motivated Kantorovich to consider another modeling
of this problem in~\cite{kantorovich}.

\begin{figure}[htb]
 \centering
 \psfrag{a}{$A$}
 \psfrag{b}{$B$}
 \psfrag{c}{$C$}
 \psfrag{d}{$D$}
 \psfrag{MT}{MTs}
 \psfrag{BS}{BSs}
 \includegraphics[width=9cm,height=6cm]{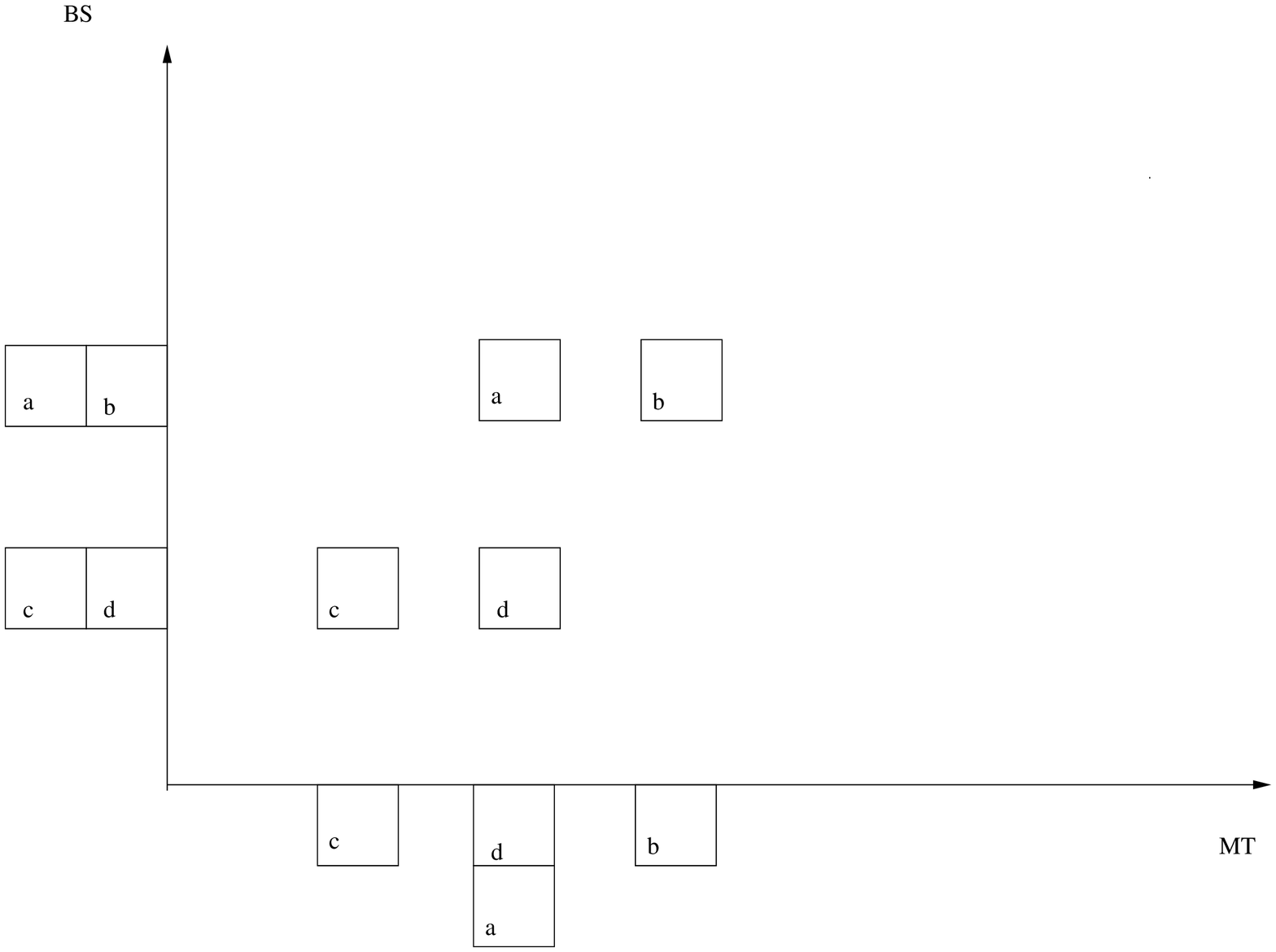}
 \caption{Kantorovich considered ``graphs'' where the projection in the first axis coincide with the mobile terminal position ($\mathrm{MT}_1=3.5$, $\mathrm{MT}_2=5$
 and $\mathrm{MT}_3=6.5$) and
 the second axis coincides with the base station position ($\BS_1=4$ and $\BS_2=6$). % These ``graphs'' are called transport plans.
 }
 \label{transport-plan-2}
 \end{figure}

Kantorovich noticed that the problem of transportation
from one location to another can be seen as
%  relaxed this problem by noticing that
 ``graphs of functions'' (called transport plans) %maps %can be viewed as positive measures
 in the product space (See Fig.~\ref{transport-plan-2}).

The idea is to minimize the objective function over the space of graphs %associate to each transport map % $f$
 %$T$~the function %measure
$g=(\mathrm{Id}\times T)$ %\#\mu$
in the product space. %, where $\mathrm{Id}$ is the identity.
Then
with the condition that each mobile terminal satisfies its uplink demand and
that the information is received at the base stations, Kantorovich's problem reads
\begin{equation}
\operatorname*{Min}_{g\in\Pi(\mu,\nu)}
\int\!\!\!\!\int_{\Omega\times\Omega} c(x,y)\,dg(x,y),\\
\end{equation}
where
\begin{equation}
\Pi(\mu,\nu)=\{g\,:\,\pi_1\#g=\mu\!\!\!\quad\text{and}\!\!\!\quad\pi_2\#g=\nu\},
\end{equation}
 is denoted the ensemble of transport plans~$g$,
 %($x$ and $y$ in $\bar{\Omega}$).
 % \begin{equation*}
 % \pi_1\#(g)=\mu\quad\text{and}\quad\pi_2\#(g)=\nu,
 % \end{equation*}
 %where %$(\pi_1(x,y),\pi_2(x,y))=(x,y)$, {\it i.e.},
 $\pi_1(x,y)$ stands for the projection on the first axis~$x$, and
 $\pi_2(x,y)$ stands for the projection on the second axis~$y$.

 %Moreover,
 %\[
 %\int_{\Omega\times\Omega} c(x,y)\,dg=\int_\Omega c(x,T(x)) f(x)\,dx.
 %\]

 The relationship between Monge and Kantorovich problems is that every
 transport map~$T$ of Monge's problem determines a transport plan %function %transport plan~
 $g_T=(\mathrm{Id}\times T)\#\mu$ in Kantorovich's problem %product space,
 with the same cost (where $\mathrm{Id}$ denotes the identity).
 However, Kantorovich's problem considers more functions than the ones coming from Monge's problem (which
can always be viewed as the product of the identity and the map~$T$), so
 we can choose from a bigger set $\Pi(\mu,\nu)$. %, denoted the class of transport plans, defined as
 %\[
 %\Pi(\mu,\nu)=\{g\text{ such that }\,\pi_1\#g=\mu\!\!\!\quad\text{and}\!\!\!\quad\pi_2\#g=\nu\}
 %\]
 %\[
 %\Pi(\mu,\nu)=\{g\in P(X\times X),\,\pi_1\#g=\mu,\pi_2\#g=\nu\},
 %\]
 %where $P(X\times X)$ is the set of probability measures in the product space~$X\times X$.

Then, every solution of Kantorovich's problem
is a lower bound to Monge's problem, {\it i.e.},
\begin{gather}
\operatorname*{Min}_{g\in\Pi(\mu,\nu)}
\int\!\!\!\!\int_{\Omega\times\Omega} c(x,y)\,dg(x,y)
\le\operatorname*{Min}_{T\#\mu=\nu}\int_\Omega c(\lvert x-T(x)\rvert)\,f(x)\,dx.%d\mu^+(x)
\end{gather}

 %where~$\Pi(\mu,\nu)$ is the class of transport plans given by
 %\[
 %\Pi(\mu,\nu)=\{g\in P(X\times X),\,\pi_1\#g=\mu,\pi_2\#g=\nu\}.
 %\]

 %Kantorovich relaxation problem (1942) is then written as
 %\begin{equation*}
 %\operatorname*{Min}_{g\in\Pi(\mu,\nu)}
 %\int_{\Omega\times\Omega} c(x,y)\,dg(x,y)
 %\end{equation*}
 %in the class of transport plans
 %\[
 %\Pi(\mu,\nu)=\{g\in P(X\times X),\,\pi_1\#g=\mu,\pi_2\#g=\nu\}.
 %\]

% %%%%%%%%%%%%%%%% Added by A. S. %%%%%%%%%%%%%%%%%%%%%%%%%%%%%%%%
%
% We denote when it exists
% \[
% M_p(\mu,\nu):=\left(\operatorname*{Min}_{T\#\mu=\nu}\int_\Omega \lvert x-T(x)\rvert^p\,f(x)\,dx\right)^{1/p}
% \]
% \[\text{and}\quad
% W_p(\mu,\nu):=\left(\operatorname*{Min}_{g\in\Pi(\mu,\nu)}
% \int\!\!\!\!\int_{\Omega\times\Omega} \lvert x-y\rvert^p\,dg(x,y)\right)^{1/p}\!\!\!\!\!\!\!.
% \]
% We are now ready to give a result on existence and uniqueness of the transport plan.
%
% \vskip 0.1cm
%
\begin{theorem}\label{theo:existenceuniqueness}
Consider the cost function~$c(\lvert x-y\rvert)=\lvert x-y\rvert^p$.
Let $\mu$ and $\nu$ be probability measures in~$\Omega$
and fix $p\geq 1$.
We assume that $\mu$ can be written\footnote{The exact condition is that $\mu$ is absolutely continuous with respect to
the Lebesgue measure.
A probability measure~$\mu$ is absolutely continuous with respect to
the Lebesgue measure if the function $F(x)=\mu((-\infty,x])$
is locally an absolutely continuous real function.
A function~$f$ is an absolutely continuous real function
if there exists an integrable function~$g$ such that
$f(x)=f(a)+\int_a^x g(t)\,dt$
} as
%\[
$d\mu=f(x)\,dx$.
%\]
% is absolutely continuous with respect to
%the Lebesgue measure $\mathcal{L}^d$ with $d=1, 2,$~or~$3$, for practical purposes.
Then the optimal value of Monge's problem coincides with the optimal value of Kantorovich's problem,
{\it i.e.}, \mbox{$M_p(\mu,\nu)=W_p(\mu,\nu)$}
and there exists an optimal
transport map from $\mu$ to $\nu$, which is also unique almost everywhere %$f$-a.e.
if~$p>1$.
\end{theorem}

This result is very difficult to obtain and it has been proved only recently (see \cite{brenier} for the case $p=2$,
and the references at \cite{villani} for the other cases).

The case that we are interested in can be characterized because the image
of the transport plan is a discrete finite set.

Thanks to optimal transport theory we are able to characterize the partitions considering general settings.
To this purpose, consider locations $(x_1,y_1)\ldots,(x_K,y_K)$, the Euclidean distance $d_i(x,y)=\sqrt{(x-x_i)^2+(y-y_i)^2}$,
and $F$ a continuous function.

\vskip 0.1cm

\begin{theorem}\label{multiplication}
Consider the problem
\begin{gather}
\mathrm{(P1)}\quad\operatorname*{Min}_{C_i} \sum_{i=1}^K \int\!\!\!\!\int_{C_i}\left[ F(d_i(x,y))+s_i\left(\int\!\!\!\!\int_{C_i}\lambda(\omega,z)\,d\omega\,dz\right)\right]\lambda(x,y)\,dx\,\,dy,
\end{gather}
\noindent where~$C_i$ is the cell partition of~$\Omega$.
Suppose that~$s_i$ are continuously differentiable, non-decreasing, and convex functions.
The problem~$\mathrm{(P1)}$ admits a solution that verifies
\begin{equation}
\mathrm{(S1)}\left\{
\begin{array}{ll}
C_i=&\big\{(x,y) : F(d_i(x,y))+s_i(N_i)+N_i\cdot s'_i(N_i)\\
&\leq F(d_j(x,y))+s_j(N_j)+N_j\cdot s'_j(N_j)\big\}\\
N_i=&\int\!\!\!\int_{C_i}\lambda(\omega,z)\,d\omega\,dz.
\end{array}
\right.
\end{equation}
\end{theorem}

\vskip 0.1cm

{\bf Proof.-} See Appendix A

\vskip 0.1cm

\begin{theorem}\label{sum}
Consider the problem
\begin{equation}
\mathrm{(P2)}\quad\operatorname*{Min}_{C_i} \sum_{i=1}^K \int\!\!\!\!\int_{C_i} \left[F(d_i(x,y))\cdot m_i\left(\int\!\!\!\!\int_{C_i}\lambda(\omega,z)\,d\omega\,dz
\right)\right]
\lambda(x,y)\,dx\,dy,
\end{equation}
\noindent where~$C_i$ is the cell partition of~$\Omega$. Suppose that~$m_i$ are derivable.
The problem~$\mathrm{(P2)}$ admits a solution that verifies
\begin{equation}
\mathrm{(S2)}\left\{
\begin{array}{ll}
C_i=&\{(x,y)\,:\,m_i(N_i) F(d_i(x,y))\,\lambda(x,y)+U_i(x,y)\\
  &\leq m_j(N_j) F(d_j(x,y))\,\lambda(x,y)+U_j(x,y)\}\\
U_i=&m'_i(N_i)\int\!\!\!\int_{C_i}F(d_i(x,y))\lambda(x,y)\,dx\,dy\\
N_i=&\int\!\!\!\int_{C_i} \lambda(\omega,z)\,d\omega\,dz.
\end{array}
\right.
\end{equation}
\end{theorem}

\vskip 0.1cm

{\bf Proof.-} See Appendix B

\vskip 0.1cm

\begin{figure}
\centering
\includegraphics[width=10cm,height=7cm]{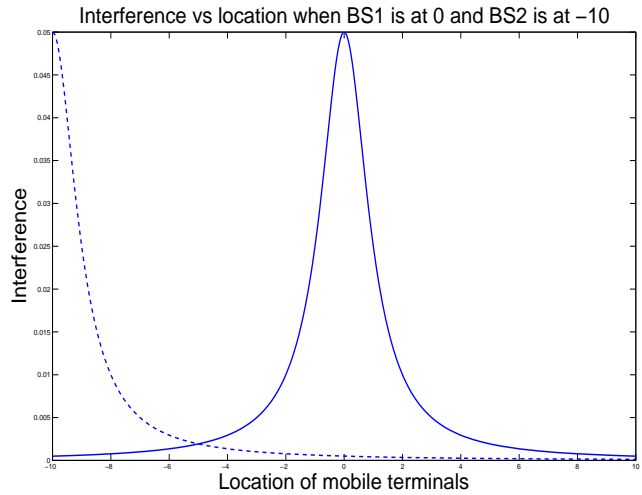}\\
\label{transportplan}
\caption{Interference as a function of location of mobile terminals when $\BS_1$
is at position $0$ (solid line) and $\BS_2$ at~$-10$ (dotted line).}
\end{figure}
%  %
%  %
%  %
%  %
\begin{figure}
\centering
\includegraphics[width=10cm,height=7cm]{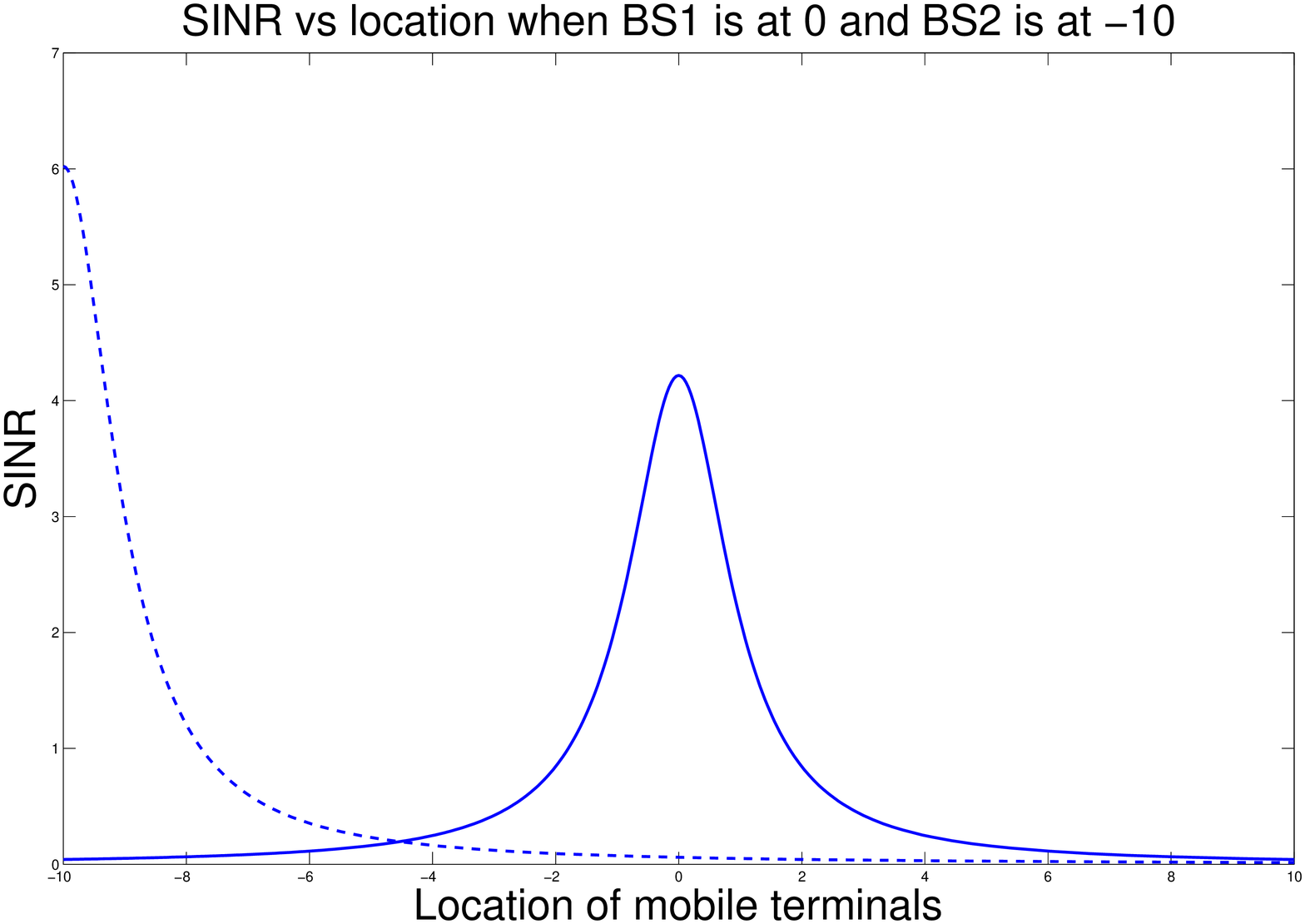}\\ \label{SINRvsLocation}
\caption{$\SINR$  as a function of location of mobile terminals when $\BS_1$ is at position $0$ (solid line)
and $\BS_2$ at~$-10$ (dotted line).}
\end{figure}
%  %
%  %  \begin{figure}
%  %  \centering
%  %  \includegraphics[width=10cm,height=7cm]{./Figures/ZoomSINRvsLocation.eps}
%  %  \caption{Zoom of the $\SINR$  as a function of the location of mobile terminals when $\BS_1$ is at position $0$ (solid line)
%  %  and $\BS_2$ is at position $-10$ (dotted line). The best equilibria is $\mathrm{eq}_1=-4.68$ with~$\SINR$ value of~$0.0025$.
%  %  }\label{ZoomSINRvsLocation}
%  %  \end{figure}
%  %
\begin{figure}
\centering
%   \subfloat[2 BSs]{\label{1D-2BS-BE}
\includegraphics[width=7cm,height=7cm]{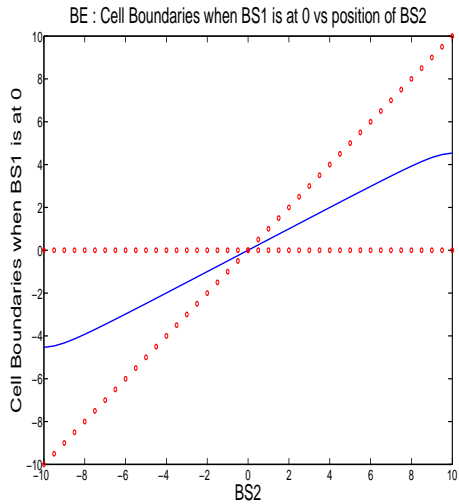}%}
\caption{Wardrop Equilibrium: Thresholds determining the cell boundaries as a function of the location of the base stations.
The network is deployed over the interval $[-10,10]$ (one-dimensional case) presented here vertically. We consider a uniform distribution of MTs
 and  we find the threshold (solid line) determining the cell boundaries
    as a function of the base stations positions (dotted lines) by changing the position of one of them.
$\BS_1$ is fixed at position~$0$ and we change the position of $\BS_2$ from $-10$ to $+10$.
}
\label{1D-2BS-BE}
\end{figure}
%  %
%  %
\begin{figure}
\centering
\includegraphics[width=10cm,height=7cm]{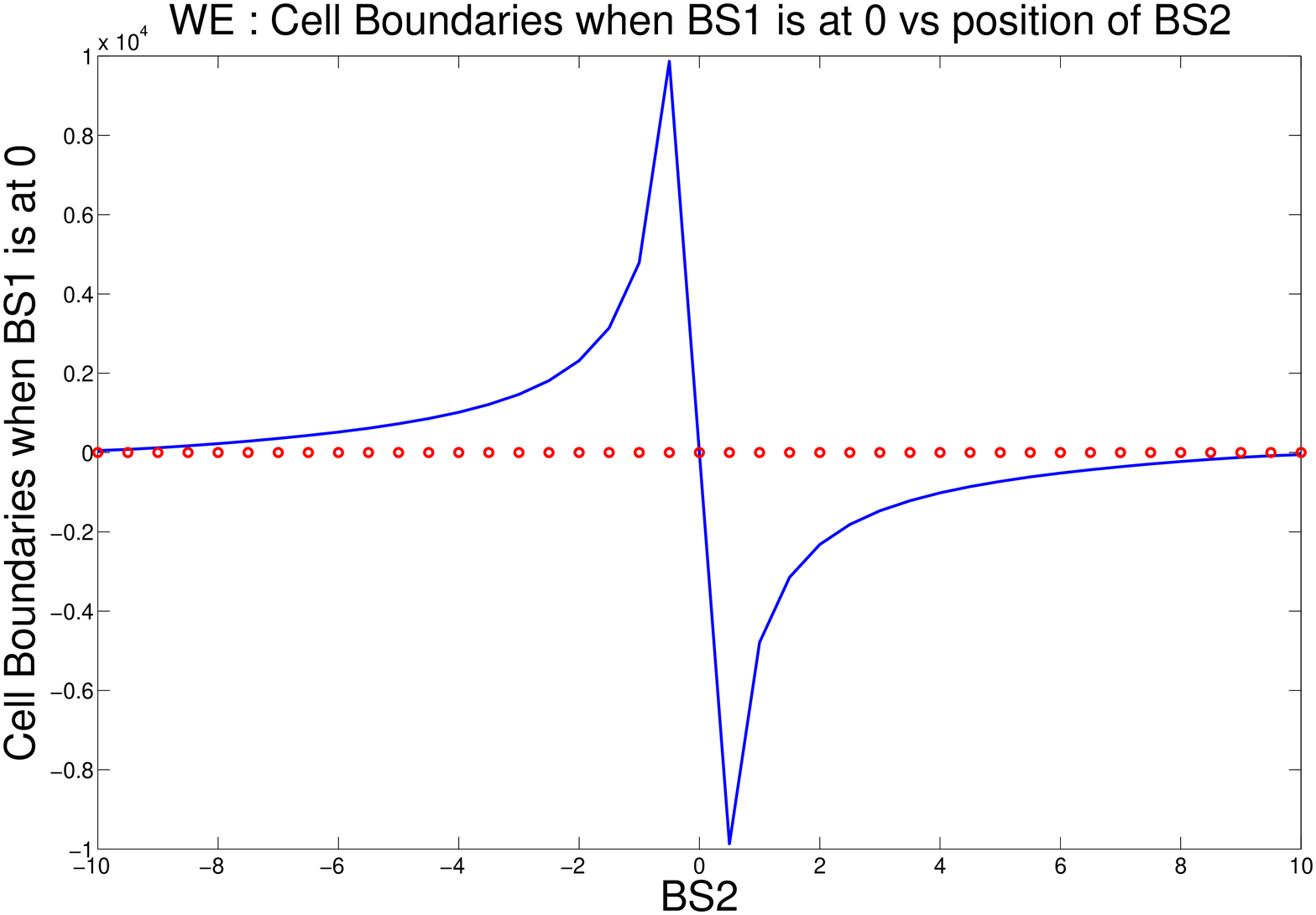}\\
\caption{Worst Equilibrium: Thresholds determining the cell boundaries (vertical axis) that give the worst equilibrium in terms of the SINR as a function of the location of~$\BS_2$ for $\BS_1$ at position~$0$.}
\label{1D-2BS-WE}
\end{figure}

Notice that in problem~$\mathrm{(P1)}$ if the functions~$s_i\equiv 0$
the solution of the system~$\mathrm{(S1)}$ becomes the well known Voronoi cells.
In problem~$\mathrm{(P2)}$ if we have that the functions~$h_i\equiv1$
we find again the Voronoi cells.
In general however, the Voronoi configuration is far from being optimal.

\section{Round Robin Scheduling Policy}\label{sectionroundrobin}

We assume that a service provider wants to minimize the total power of the network
while maintaining a certain average throughput of~$\theta$ to each mobile terminal of the system
using the round robin scheduling policy given by problem
\begin{equation}\label{roundrobin}
(\mathrm{RR})\quad\operatorname*{Min}_{C_i}\sum_{i=1}^K\int\!\!\!\!\int_{C_i}\sigma^2 (R^2+d_i(x,y)^2)^{\xi/2}(2^{N_i\theta}-1)\lambda(x,y)\,dx\,dy.
\end{equation}

We see that this problem is an optimal transportation problem like the one in~$\mathrm{(P1)}$
with cost function given by
\begin{gather}
F(d_i(x,y))=\sigma^2 (R^2+d_i(x,y)^2)^{\xi/2}%(2^{N_i\theta(x,y)}-1)
\quad\text{and}\\
m_i(x,y)=(2^{N_i\theta}-1).
\end{gather}
From the previous theorem, we can derive an explicit expression
for this configuration.

{\bf Proposition.-} There exist a unique optimum given by
\begin{align}
C_i =&\Big\{x_0\in\Omega\ :d_i(x_0,y_0)^p+h_i(N_i)+N_i h_i'(N_i)\nonumber\\
 &\leq d_j(x_0,y_0)^p+k_j(N_j)+N_j k_j'(N_j),\, \forall j\neq i\Big\}\\
N_i =&\int\!\!\!\int_{C_i} \lambda(x_0,y_0)\,dx_0\,dy_0.
\end{align}

% {\bf Proof.-} See Appendix~A.

%\begin{itemize}
%\item If $h_i$ are lower semi-continuous functions, then there exists an optimum.
%\item If $h_i$ are lower semi-continuous functions and $\eta_i:=t h_i(t)$
% are strictly convex, then there exists a unique optimum.
% \item If $h_i$ are differentiable in $]0,1]$ and continuous in $0$
% and $\eta_i$ are convex, then there exists a unique optimum.
% \end{itemize}
% Defining
% \[
% (*)\left\{
% \begin{array}{ll}
% A_i =&\{x\in\Omega\ :\\ &\lvert x-x_i\rvert^p+h_i(c_i)+c_i h_i'(c_i)
% <\lvert x-x_j\rvert^p+h_j(c_j)+c_j h_j'(c_j)\quad\forall j\neq i\}\\
% c_i=&\int_{A_i} f(x)\ dx
% \end{array}\right.
% \]
% \begin{itemize}
% \item If $h_i$ are differentiable and continuous in $0$,
% then $(*)$ is a necessary optimality condition.
% \item If $h_i$ are differentiable in $]0,1]$ and continuous
% in $0$ and $\eta_i$ are convex, then $(*)$ is a
% necessary and sufficient optimality condition.
% \end{itemize}

%%%%%%%%%%%%%%%%%%%%%%%%%%%%%%%%%%%%%%%%%%%%%%%%%%%%%%%%%
%
% Stochastic Approximation for a certain underlying dynamics
% -> Rest points
% ->
% \begin{gather}
% \min\sum_s \int_0^{x_s} f_s(z)dz\\
% R x\leq c\\
% 	x\geq 0
% \end{gather}
%
% {\bf Discrete case vs continuous case}
%
% Let's consider $\mu^+=\sum_i a_i\delta_{x_i}$
% and $\mu^-=\sum_j b_j\delta_{y_j}$
% both atomic measures of equal total mass.
% If we assume $\{x_i\}$ and $\{y_j\}$
% are atomic measures of equal total mass.
%
% If $\mu^+=$

Let's see a direct application of our results:
\begin{example}
Consider a network of~$N=2500$ mobile terminals %the case when the number of mobile terminals in the network is given by~$N=2500$,
distributed according to~$\lambda(x)$ in $[0,L]$
(for example, with $L=5.6$ miles for WiMAX radius cell). %he domain~$[0,1]$ and
We consider two base stations at position~$\BS_1=0$
and~$\BS_2=L$ and~\mbox{$R=1$}. % (for example $R=10$ meters).
Then, the mobile association threshold (the boundary between both cells, i.e.,
the location at which the mobile terminals obtain the same throughput by connecting to
any of both base stations)
is reduced to find~$x$ such that the following equality holds:
\begin{gather}
(2^{N_1\theta}-1)(1+x^2)\lambda(x)+2^{N_1\theta}\theta\log 2 \left[x+\frac{x^3}{3}\right]
=\nonumber\\(2^{N_2\theta}-1)(1+(1-x)^2)\lambda(x)+
2^{N_2\theta}\theta\log 2 \left[\frac{4}{3}-2x+x^2-\frac{x^3}{3}\right]
\end{gather}
Notice that this is a fixed point equation on~$x$.
If the mobile terminals are distributed uniformly, the optimal solution is given by~$C_1=[0,1/2~L)$
%~$x=1/2$
and~$C_2=[1/2~L,L]$,
which is the solution that Voronoi cells would give us and in that case the number of mobile terminals
connected to each base station would be given by
\begin{equation}
N_1=N_2=1250.
\end{equation}
However, if the deployment distribution of the mobile terminals is more concentrated near $\BS_2$ than $\BS_1$,
consider for example~$\lambda(x)=2x$,
the optimal solution is given by \mbox{$C_1=[0,q)$} and \mbox{$C_2=[q,L]$} with $q=0.6027~L$ and
\begin{equation}
N_1=908\quad\textrm{and}\quad N_2=1592.
\end{equation}
Notice that in the global optimization solution, the number of mobile terminals
connected to $\BS_1$ is smaller that the number of mobile
terminals connected to $\BS_2$. However,
the cell size is bigger.

 %~$N_2=0.6367$.
%(see Fig.~\ref{distribution2x}).
\end{example}
 \begin{figure}[htp]
 \centering
 \subfloat[Equilibrium in Non-Uniform distribution]{\label{distribution2x}
 \includegraphics[width=7cm,height=7cm]{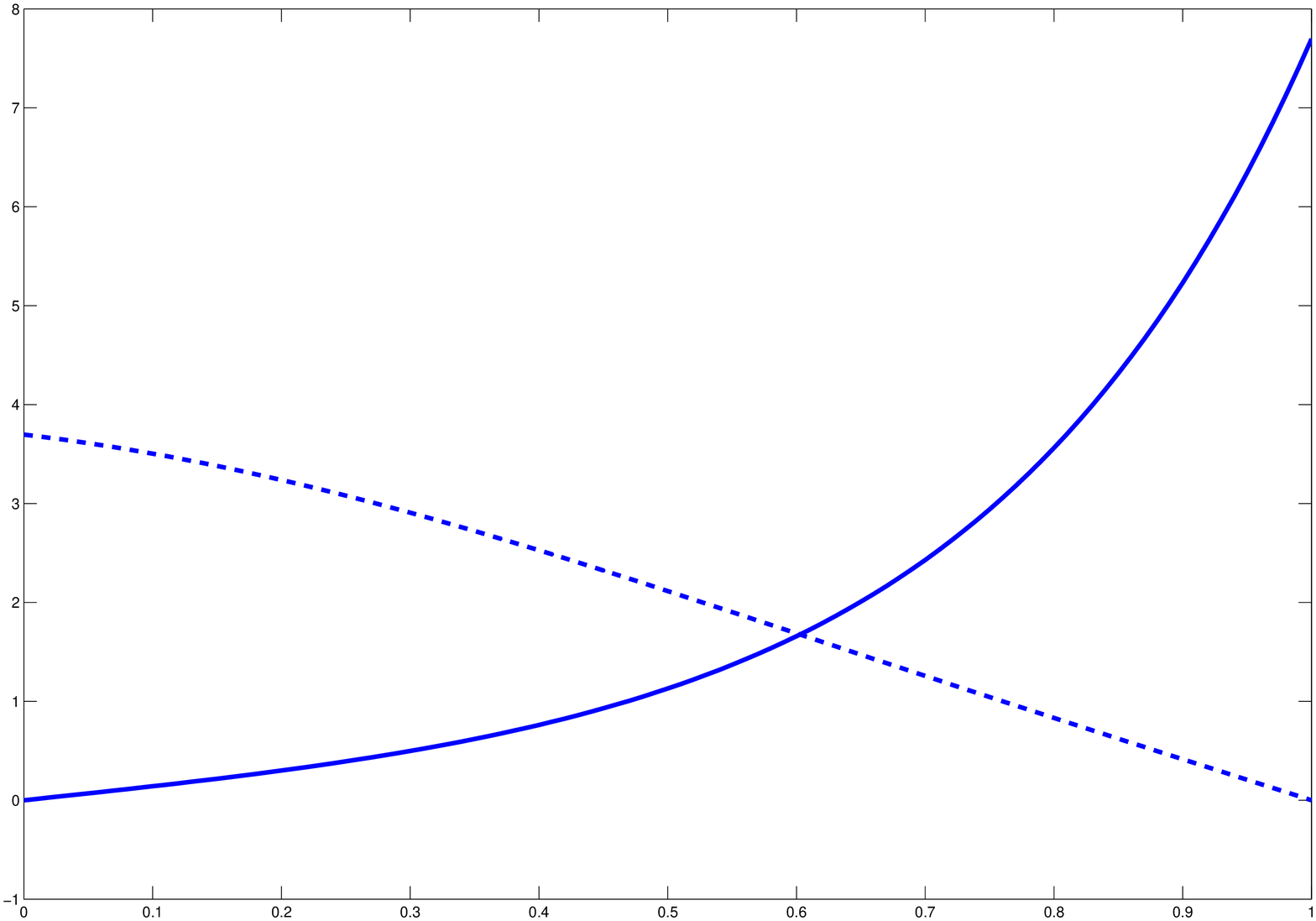}
 }
  \caption{
  Thresholds determining the cell boundaries (vertical axis) of the best equilibrium in terms of the SINR
  as a function of the location of~$BS2$ for BS1 at position~$0$ when we consider a non-homogeneous distribution
  given by $\lambda(x)=(L-x)/2L^2$.
  Example: equilibrium when then distribution of mobile terminals is given by $\lambda(x)=2x$ in the interval~$[0,L]$
  and the positions of the base stations are~$\BS_1=0$ and~$\BS_2=L$.}
  \label{distribution2x}
 \end{figure}
\begin{figure}
\centering
\includegraphics[width=7cm,height=7cm]{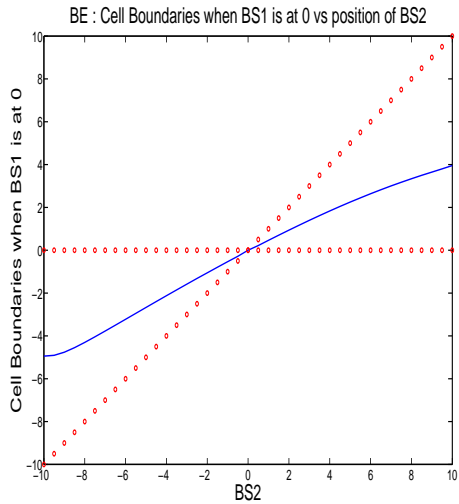}
\caption{Wardrop Equilibrium in the Non-homogeneous case:
The network is deployed over the interval~$[-10,10]$ (one-dimensional case) presented here vertically.
We find the threshold (solid line) determining the cell boundaries
as a function of the base stations positions (dotted lines) by changing the position of one of them.
$\BS_1$ is fixed at position~$0$ and we change the position of $\BS_2$ from $-10$ to $+10$.
The deployment distribution of the MTs is given by $\lambda(x)=(L-x)/2L^2$.
}
\label{1D-Non-homogeneous-BE}
\end{figure}

%\subsection{Penalization function}\label{subsec:penaliza}
%
%Notice that the penalization function or the case when the number of
%mobile terminals is greater than the number of carriers available in
%the cell is the mixture of both problems.
%

%%%%%%%%%%%%%%%%%% FAIRNESS PROBLEM %%%%%%%%%%%%%%%%%%%%%%%%%%%%%%%%%%%%%%%%%%%%%%%%%%
%
\section{Fairness problem}\label{fairness}

As we mention in section~\ref{model} the solution given by previous section~\ref{sectionroundrobin}
is optimal but may not be fair to all the mobile terminals since it will give higher throughput to the
mobile terminals that are near the base stations.
To deal with this problem we considered the fairness problem given by
\begin{gather*}
\operatorname*{Min}\sum_{i=1}^K\int\!\!\!\!\int_{C_i}
\frac{1}{\alpha-1}\left(\sigma^2 (R^2+d^2_i(x,y))^{\xi/2}\right)^{\alpha-1}
(2^{N_i\theta}-1)^{\alpha-1}\lambda(x,y)\,dx\,dy.
\end{gather*}
As we can see this is also an optimal transportation problem~$\mathrm{(P1)}$
where the functions considered in this setting are given by
\begin{equation}
F(d_i(x,y))=\frac{1}{\alpha-1}\left(\sigma^2 (R^2+d^2_i(x,y))^{\xi/2}\right)^{\alpha-1}
\end{equation}
\begin{equation}
m_i(x,y)=(2^{N_i\theta}-1)^{\alpha-1}.
\end{equation}
Using Theorem~\ref{multiplication} we are able to characterize the optimal cells for
any $\alpha$ considered.

\section{Rate fair allocation policy}\label{sectionratefair}

In this framework, we give the possibility to mobile terminals to associate to the base station
they prefer in order to minimize their power cost function while maintaining, as quality of service
measurement, an average throughput of~$\theta$.

As we presented in Section~\ref{model}, this problem is equivalent to
\begin{equation}
(\mathrm{RF})\quad\operatorname*{Min}_{C_i}\sum_{i=1}^K\int\!\!\!\!\int_{C_i}\sigma^2(R^2+d^2_i(x,y))^{\xi/2}\lambda(x,y)\,dx\,dy.
\end{equation}

Notice that this problem is equivalent to $\mathrm{(P1)}$ where the functions
\begin{equation}
F(d_i(x,y))=\left(\sigma^2 (R^2+d_i(x,y)^2)^{\xi/2}\right),
\end{equation}
and $s_i\equiv 1$.
The problem has then a solution given by

{\bf Proposition.-} There exist a unique optimum given by
\begin{equation}
\begin{array}{ll}
C_i =&\Big\{x\in\Omega:
\sigma^2(R^2+d^2_i(x_0,y_0))^{\xi/2}\\
&\leq \sigma^2(R^2+d^2_j(x_0,y_0))^{\xi/2},\forall j\neq i\Big\}\\
N_i =&\int\!\!\!\int_{C_i} \lambda(x_0,y_0)\,dx_0\,dy_0,
\end{array}
\end{equation}
which is represented by the Voronoi cells.

\subsection{Penalization function}\label{subsec:penalization}

Notice that the penalization function or the case when the number of
mobile terminals is greater than the number of carriers available in
the cell in the rate fair allocation policy case
is equivalent to $\mathrm{(P1)}$ where the functions

\begin{equation}
F(d_i(x,y))=\left(\sigma^2 (R^2+d_i(x,y)^2)^{\xi/2}\right).
\end{equation}

\begin{equation}
s_i(N_i)=
\kappa_i(N_i)=
\left\{
\begin{array}{cl}
0&\textrm{ if }N_i\leq\mathrm{MAX}_i,\\
\bar\kappa_i(N_i-\mathrm{MAX}_i)&\textrm{ if }N_i> \mathrm{MAX}_i.
\end{array}
\right.
\end{equation}

The problem has then a solution given by

{\bf Proposition.-} There exist a unique optimum given by
\begin{equation}
\left\{
\begin{array}{ll}
C_i=&\big\{(x,y) : \sigma^2(R^2+d^2_i(x_0,y_0))^{\xi/2}+s_i(N_i)+N_i\cdot s'_i(N_i)\\
&\leq \sigma^2(R^2+d^2_i(x_0,y_0))^{\xi/2}+s_j(N_j)+N_j\cdot s'_j(N_j)\big\}\\
N_i=&\int\!\!\!\int_{C_i}\lambda(\omega,z)\,d\omega\,dz.
\end{array}
\right.
\end{equation}

% \section{Uplink Case}
%
%
%
%
%
% {\it Penalization function}
% As an illustration example, suppose that on the network~$\Omega=[0,1]$
% there are two base stations at coordinates~\mbox{$x_1=1/4$} and~\mbox{$x_2=3/4$}.
% Assume that mobile terminals are uniformly distributed, and consider the case when~$\xi=2$.
%
% Suppose the first base station can handle more downlink demand than the second one,
% as for example the first base station uses a IEEE 802.16 (WiMaX) technology while the second one uses UMTS technology,
% so that the penalization cost are
% \[
% h_1(t)=t\quad\text{and}\quad h_2(t)=(1+\varepsilon)t.
% \]
%
% Then the optimum cell configuration~$(C^*_1,C^*_2)$ is given by
% \[
% C^*_1=[0,\lambda_{\varepsilon}^*[,\quad C^*_2=]\lambda_{\varepsilon}^*,1]\quad\text{~with}\quad
% \lambda_{\varepsilon}^*=\frac{1}{2}+\frac{\varepsilon}{5+2\varepsilon},
% \]
% whereas the equilibrium cell configuration~$(C^E_1,C^E_2)$ will be
% $
% C^E_1=[0,\lambda_{\varepsilon}^E[,\quad C^E_2=]\lambda_{\varepsilon}^E,1]$~with
% \[
% \lambda_{\varepsilon}^E=\frac{1}{2}+\frac{\varepsilon}{6+2\varepsilon}\leq\lambda_{\varepsilon}^*.
% \]
%

%%%%%%%%%%%%%%%% Commented by AS
%{\it Example 2.-}
%
%%%%%%%%%%%%%%%%%%%%%%%%%%%%%%%%%%%%%%%%%%%%%%%%%%%%%%%%%%%%%%%%%%%%%%%%%%%%%%%%%%%%%%%%%%%%%%%%%%%%%%%%%%%%%%%%%%%%
\section{Performance Gap}\label{sec:PoA}
As a second example consider again the case when the mobile terminals are uniformly distributed on~$\Omega=[0,1]$
but this time the two antennas are located at coordinates~$x_1=0$ and~$x_2=1$.
Consider the case when~$p=1$ and
\[
s_1(x)=100\quad\text{and}\quad
s_2(x)=\left\{
\begin{array}{ll}
0&\quad\text{for}\quad 0\le x\le0.999\\
1&\quad\text{for}\quad 0.999\le x\le 1.
\end{array}
\right.
\]

Then the equilibrium cell configuration~$(C^E_1,C^E_2)$ is given by
\[
C^E_1=\emptyset\quad\text{and}\quad C^E_2=[0,1],
\]
and the optimum cell configuration~$(C^*_1,C^*_2)$ is
\[
C^*_1=[0,001[\quad\text{and}\quad C^*_2=]0.001,1].
\]
The optimum is very unfair for mobile terminals living in the first cell~$C^*_1$,
who pay~$x+100$, whereas the other mobile terminals just pay the distance from~$1$.
This is a toy example but it gives an idea of the performance gap between
the centralized and the decentralized scenarios, also known as Price of Anarchy.
    \begin{figure} %[htp]
    \centering
%    \subfloat[3 BSs]{\label{1D-3BS}
\includegraphics[width=7cm,height=7cm]{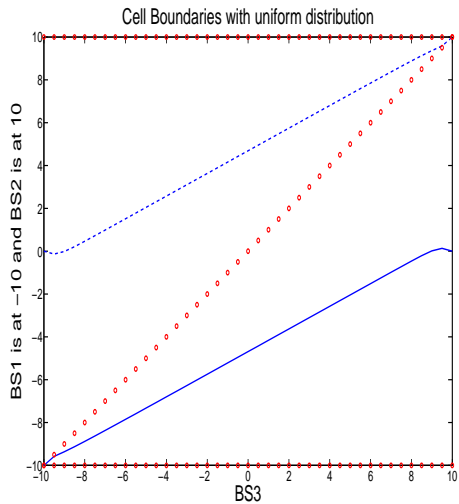}
%}
    \caption{
Wardrop Equilibrium with Multiple Base Stations:
The grid area network is the interval $[-10,10]$ presented here vertically. We consider a uniform distribution of MTs.
We find the threshold (solid and dashed lines) determining the cell boundaries
    as a function of the base stations positions (dotted lines) by changing the position of one of them.
$\BS_1$ is fixed at position~$-10$ and $\BS_2$ is fixed at position~$10$ and we change the position of $\BS_3$ from $-10$ to $+10$.
    }
    \label{1D-3BS}
    \end{figure}

\section{Numerical Simulations} \label{secnum}

\begin{figure}
\centering
%    \subfloat[Uniform distribution of users]{\label{2D-KBS}\includegraphics[width=8cm,height=7cm]{./Figures/2D-Cell-Formation.eps}}
\includegraphics[width=8cm,height=7cm]{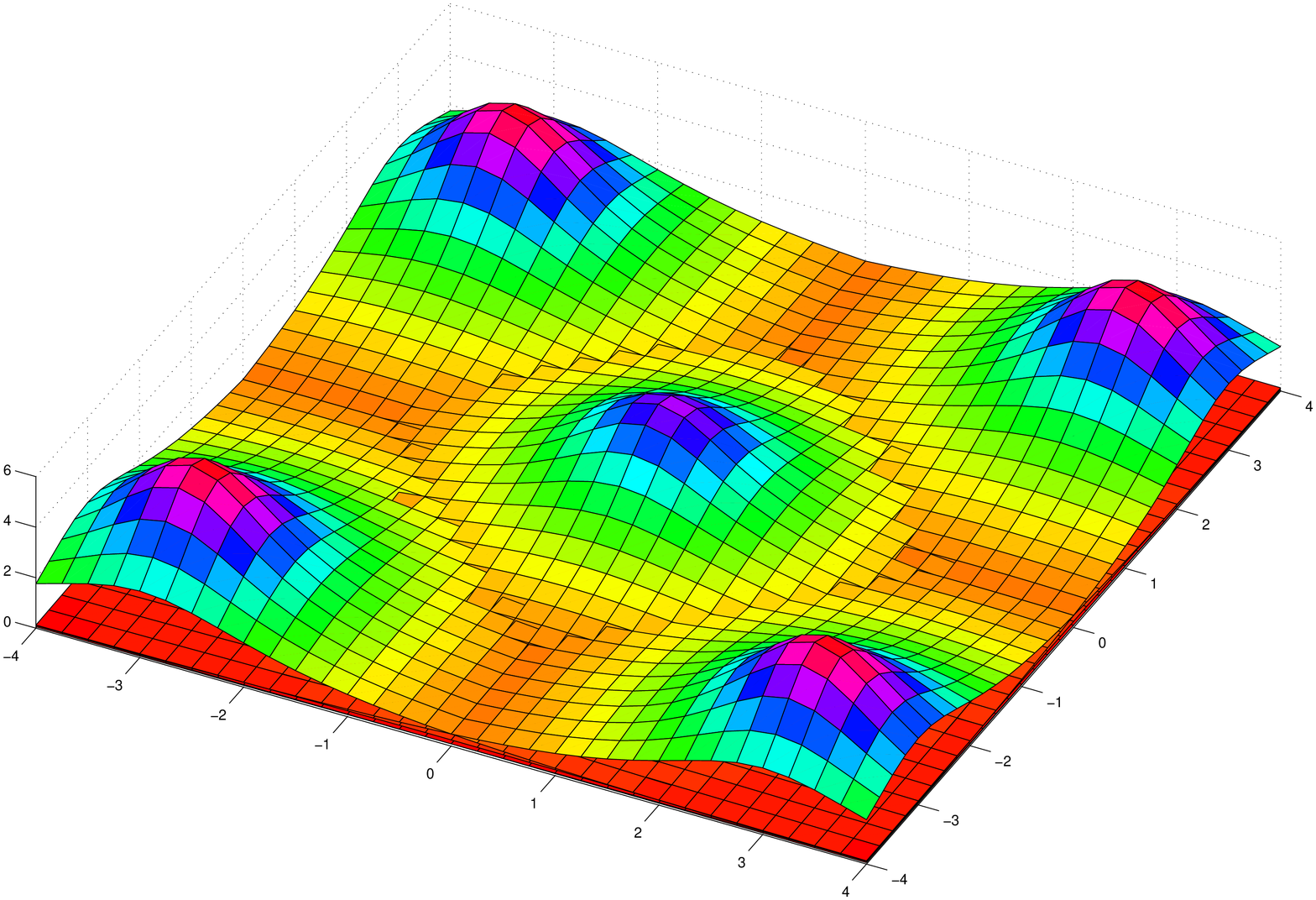}\\
\caption{Wardrop Equilibrium in the 2D case:
The grid area network is the square $[-4,4]\times[-4,4]$. We set the noise parameter $\sigma=0.3$
and we set four BSs at positions $\BS_1=(-3,-3)$ $\BS_2=(3,-3)$ $\BS_3=(3,3)$ $\BS_4=(-3,3)$ and one at the origin $\BS_5=(0,0)$.
We determine the cell boundaries (deep lines) for the uniform distribution of users}
\label{2D-KBS}
\end{figure}
%  %
\begin{figure}
\centering
\includegraphics[width=10cm,height=7cm]{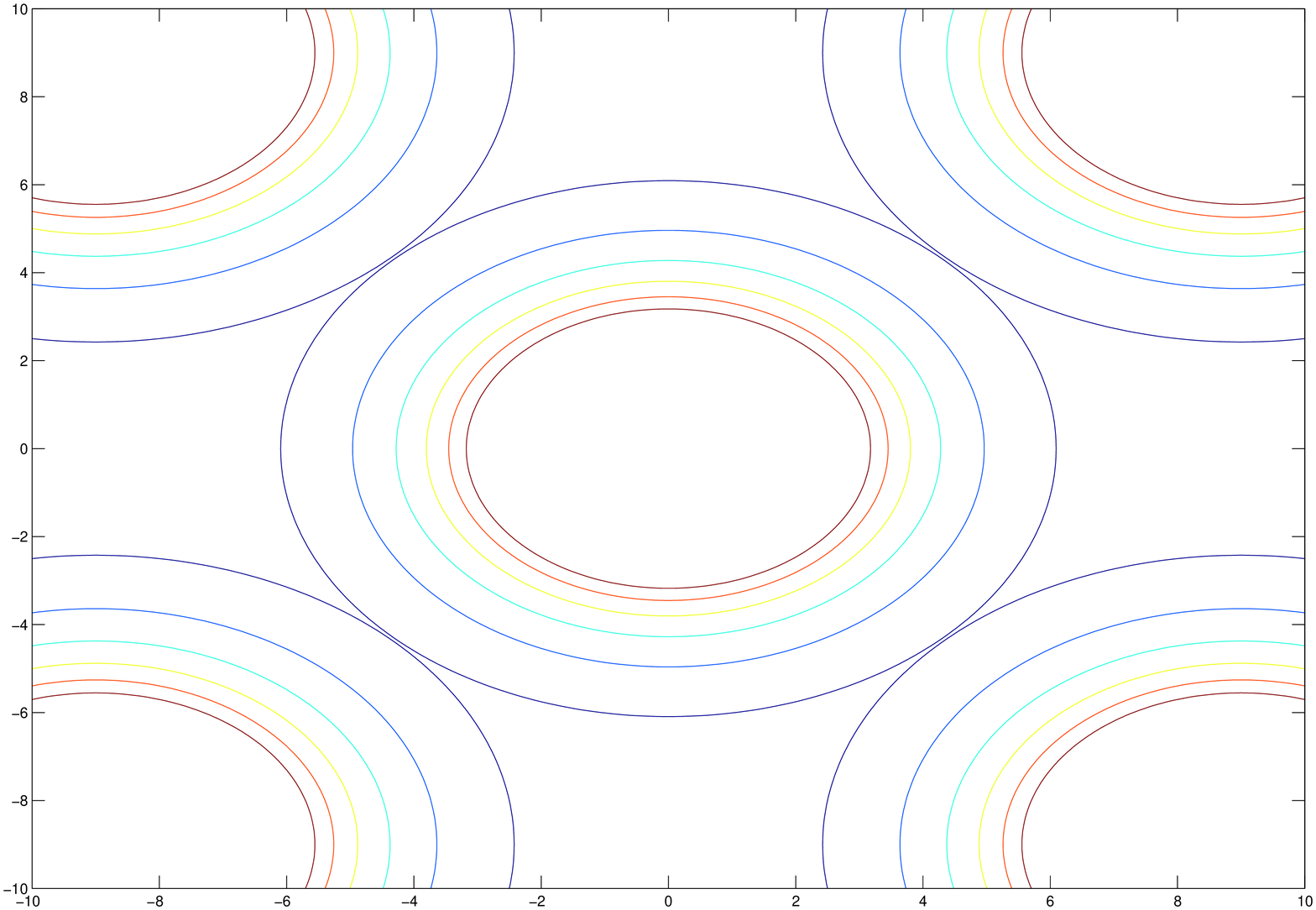}\\
\caption{2D case: Cell contours of the best equilibrium with uniform distribution of users.}
\label{2D-KBS}
\end{figure}
%  %
\begin{figure}
\centering
%      \subfloat[Non-uniform distribution of users]{\label{2D-NH}
\includegraphics[width=8cm,height=7cm]{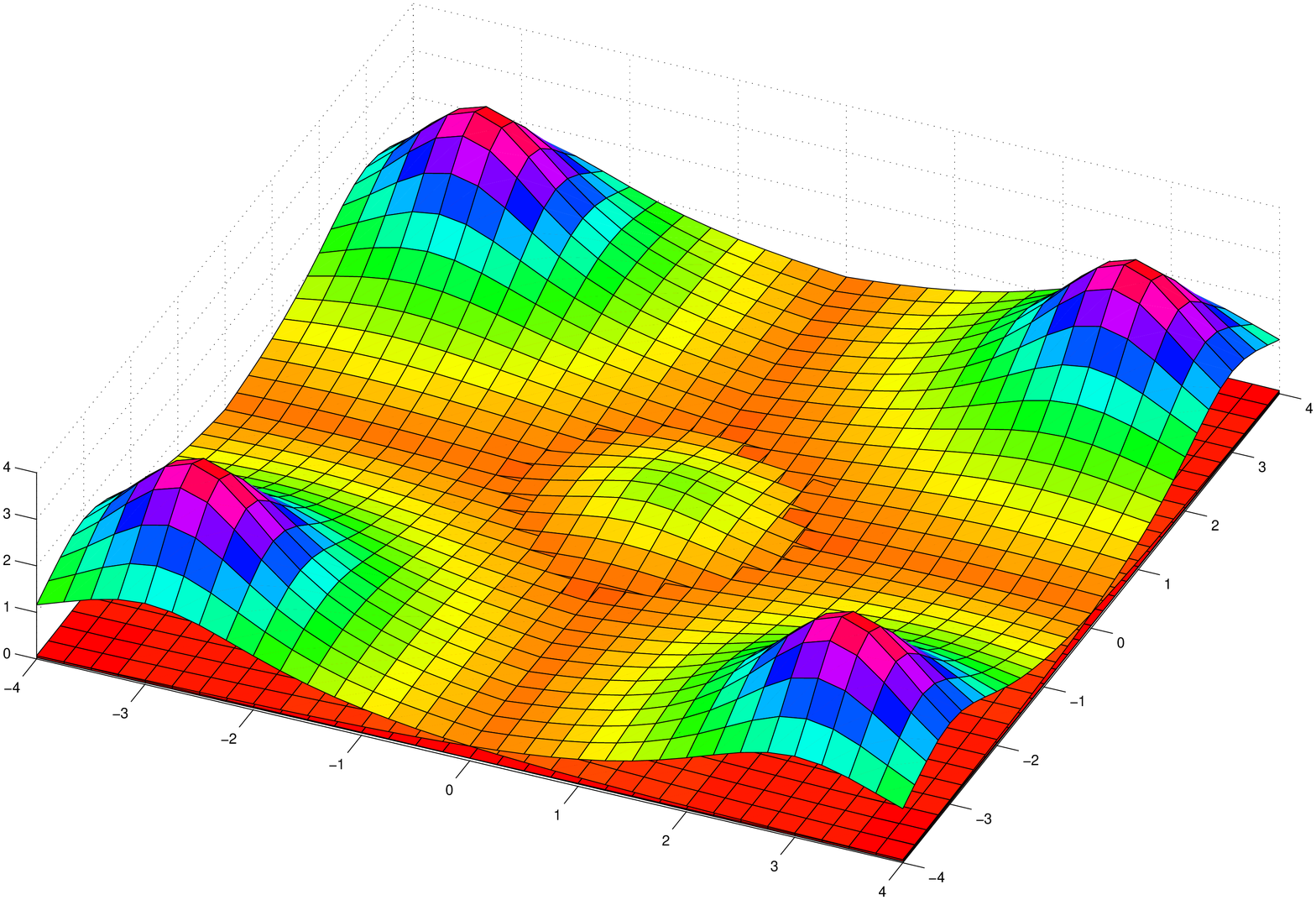}
% }
%  %  \includegraphics[width=10cm,height=7cm]{./Figures/2D-NH.eps}\\
\caption{Wardrop Equilibrium in 2D Non-Uniform Case:
The grid area network is the square $[-4,4]\times[-4,4]$. We set the noise parameter $\sigma=0.3$
and we set four BSs at positions $\BS_1=(-3,-3)$ $\BS_2=(3,-3)$ $\BS_3=(3,3)$ $\BS_4=(-3,3)$ and one at the origin $\BS_5=(0,0)$.
We determine the cell boundaries (deep lines)
for the non-uniform distribution of users given by
\mbox{$\lambda(x,y)=(L^2-(x^2+y^2))/K$} where $K$ is a normalization factor. The latter situation takes into account
when mobile terminals are more concentrated in the center
and less concentrated in suburban areas.
% We observe that the cell size of the base station
% at the center is smaller than the others at the suburban areas.
% This can be explained by the fact that as the density of
% users is more concentrated in the center the
% interference is greater in the center than in the suburban areas
% and then the $\SNR$ is smaller in the center.
% However the quantity of MTs connected to the BS at the center is greater than in the suburban areas.
}
\end{figure}

\begin{figure}
\centering
\includegraphics[width=10cm,height=7cm]{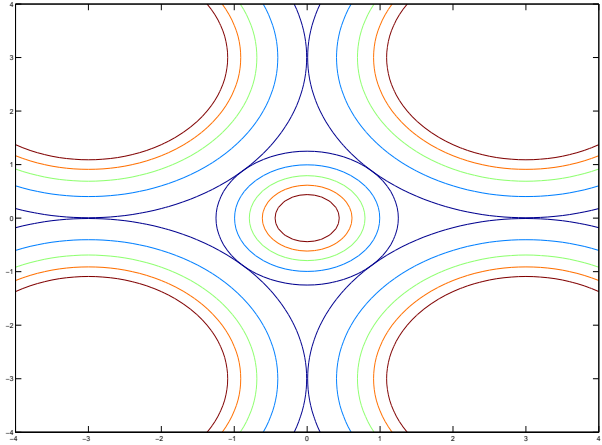}\\
\caption{2D Non-Uniform: Cell contours of the best equilibrium with non-uniform distribution of users.}
\label{2D-NH}
\end{figure}

In this section, we present several numerical results that validate our theoretical model.
\subsection{One-dimensional case}\label{A}
We first consider the one-dimensional case and we
consider a uniform distribution of users in the interval~$[-10,10]$.
We set the noise parameter~$\sigma=0.3$.
In Fig.~\ref{1D-2BS-BE},
we fix one base station~$\BS_2$ at position~$0$
and take as parameter the position of base station~$\BS_1$.
We consider as path loss exponent~$\xi=2$.
% In the $\SNR$-association game we found two pure equilibria:
% the best equilibria at position $\mathrm{eq}_1=-4.68$
% with~$\SNR$ value of~$2.5\times 10^{-3}$ %$0.0025$
% and the worst equilibria at position $\mathrm{eq}_2=78.69$ with $\SNR$ value of~$1.4769\times 10^{-9}$.
% It is known than any other mixed equilibria will give lower values of~$\SNR$ than the best equilibria.
% From now on we will be interested only in the best equilibrium.
Red lines shows the positions of the BSs.
We are able to determine the cell boundary (solid blue curve) from~$\BS_1$ and $\BS_2$
at different positions.
In Fig.~\ref{1D-3BS}, we fix two base stations~$\BS_1=-10$ and~$\BS_2=10$
and we take as parameter the position of base station~$\BS_3$.
Red lines shows the positions of the BSs.
We determine the cell boundary (solid blue curve) from~$\BS_1$ and $\BS_3$
and the cell boundary (dotted blue curve) from $\BS_2$ and $\BS_3$.

%See Fig.~\ref{1D-2BS-BE} and~\ref{1D-3BS}.
% We found that even in the one-dimensional case,
% the results of~\cite{spatial} are not-valid,
% the cells are convex and monotone inside the network.

% % \subsection{One-dimensional case: Non-uniform distribution of users}
% %
% % In this case we consider a non-uniform distribution of users
% % $\lambda(x)=(L-x)/L^2$ under the same setting as in \ref{A}.
% % we found again that the cells are convex and monotone inside the network.

\subsection{Two-dimensional case: Uniform and Non-Uniform distribution of users}
We consider the two-dimensional case.
We consider the square~$[-4,4]\times[-4,4]$ and the noise parameter~$\sigma=0.3$.
We set five base stations at positions~$\BS_1=(-3,-3)$,
$\BS_2=(3,-3)$, $\BS_3=(-3,3)$, $\BS_4=(3,3)$,
and $\BS_5=(0,0)$.
%Numerically we observe again that the cells are convex and monotone inside the domain.
We determine the cell boundaries for the uniform distribution of MTs
(see Fig.~\ref{2D-KBS}) and we compare it to the cell boundaries
for the non-uniform distribution of MTs
%
%We consider the two-dimensional case and this time we
%consider a non-uniform distribution of users in the square~$[-L,L]\times[-L,L]$
given by
\mbox{$\lambda(x,y)=(L^2-(x^2+y^2))/K$}
where $K$ is a normalization factor.
The latter situation can be interpreted as the situation
when mobile terminals are more concentrated in the center
and less concentrated in suburban areas as in Paris, New York or London.
We observe that the cell size of the base station~$\BS_5$
at the center is smaller than the others at the suburban areas.
This can be explained by the fact that as the density of
users is more concentrated in the center the
interference is greater in the center than in the suburban areas
and then the $\SINR$ is smaller in the center.
However the quantity of users is greater than in the suburban areas.
 %We consider the two-dimensional case and this time we
% % %consider a non-uniform distribution of users in the square~$[-L,L]\times[-L,L]$
% % %given by
% % %\mbox{$\lambda(x,y)=(L^2-(x^2+y^2))/K$}
% % %where $K$ is a normalization factor.
% % %This situation can be interpreted as the situation
% % %when mobile terminals are more concentrated in the center
% % %and less concentrated in suburban areas as in Paris, New York or London.
% % %We observe that the cell size of the base station~$\BS_5$
% % %at the center is smaller than the others at the suburban areas.
% % %This can be explained by the fact that as the density of
% % %users is more concentrated in the center the
% % %interference is greater in the center than in the suburban areas
% % %and then the $\SINR$ is smaller in the center.
% % %However the quantity of users is greater than in the suburban areas. %...{\bf TO COMPLETE TOMORROW}
% % %
% % %
% % %See Fig.~\ref{2D-NH}.

\section{Conclusions and Future Perspectives}\label{secconclusion}

In the present work, we have studied the mobile association problem
in the downlink scenario.
The objective is to determine the spatial locations at which 
mobile terminals would prefer to connect to a given base station rather than to other
base stations in the network if they were offered that possibility
(denoted decentralized scenario).
We are also interested in the spatial locations
which are more convenient from a centralized or from a network operator point of view.
In both approaches, the optimality depends upon the context.
In the considered cases, we consider
the minimization of the total power
of the network,
which can be considered as an energy-efficient objective,
while maintaining a certain level of throughput 
for each user connected to the network.
We have proposed a new approach using optimal transport
theory for this mobile association problems and we have been able
to characterize these mobile associations
under different policies.

The present work can be extended in several different directions.
One of these possible directions is to study the price of anarchy
between the centralized and decentralized scenario.
As we presented in Section~\ref{sec:PoA},
the considered example give us an indication
that the price of anarchy should be unbounded
but currently we don't have 
precise bounds.
The price of anarchy should be studied in both scenarios:
the sum of a function and the multiplication.
It should be interesting to study the application
in the particular case when the network
is an LTE network.
Since our model is quite simplified in order to
obtain exact solutions we could include
the cases for the fading and shadowing effects.
It is implicitly considered that
the number of users in the network is
stationary, but since at different times
of the day there are different number of users,
this management capabilities should be taken
into account.

\section*{Acknowledgements}
The authors would like to warmly thank Chlo\'e Jimenez from {\sl Universit\'e de Brest} for interesting discussions.
The last author was partially supported by Alcatel-Lucent
within the Alcatel-Lucent Chair in Flexible Radio at Sup\'elec.

\section*{Appendix A}
%\begin{theorem}\label{multiplication}
Consider the problem~$\mathrm{(P1)}$
\begin{gather}
\operatorname*{Min}_{C_i} \sum_{i=1}^K \int\!\!\!\!\int_{C_i}\left[ F(d_i(x,y))+s_i\left(\int\!\!\!\!\int_{C_i}\lambda(\omega,z)\,d\omega\,dz\right)\right]\lambda(x,y)\,dx\,\,dy,
\end{gather}
\noindent where~$C_i$ is the cell partition of~$\Omega$.
Suppose that~$s_i$ are continuously differentiable, non-decreasing, and convex functions.
The problem~$\mathrm{(P1)}$ admits a solution that verifies
\begin{equation}
\mathrm{(S1)}\left\{
\begin{array}{ll}
C_i=&\big\{(x,y) : F(d_i(x,y))+s_i(N_i)+N_i\cdot s'_i(N_i)\\
&\leq F(d_j(x,y))+s_j(N_j)+N_j\cdot s'_j(N_j)\big\}\\
N_i=&\int\!\!\!\int_{C_i}\lambda(\omega,z)\,d\omega\,dz.
\end{array}
\right.
\end{equation}
%\end{theorem}

{\bf Proof.-}
The proof is based on Proposition~3.5 of Crippa et al.~\cite{crippa}.
We include the proof for completeness and because part of it (mainly the existence of the solution) will be used in the proof of the following theorem.
Notice that we have considered the case \mbox{$F(d_i(x,y))=\lvert (x,y)-(x_i,y_i)\rvert^p$}, but this holds for any continuous function $F$.

From Section~\ref{secbasics}, let us recall that Monge's problem can be stated as follows: given two probability measures, $\mu$ and $\nu$,
and a constant $p\ge 1$ we consider the minimization problem (denoted by $M_p(\mu,\nu)$):
\begin{equation}
M_p(\mu,\nu):=\inf\left\{\left(\int_\Omega\lvert x-T(x)\rvert^p\,d\mu(x)\right)^{1/p}\,:\, T:\Omega\to\Omega\,\,\textrm{Borel and such that}\,\,T\#\mu=\nu\right\}.
\end{equation}

The relaxed formulation of Monge's problem (denoted by $W_p(\mu,\nu)$) can be stated as follows
\begin{equation}
W_p(\mu,\nu):=\inf_{\gamma\in\Pi(\mu,\nu)}\left\{\left(\int_{\Omega\times\Omega}\lvert x-y\rvert^p\,d\gamma(x,y)\right)^{1/p}\right\},
\end{equation}
where $\Pi(\mu,\nu)$
is the set of probability measures such that
$\pi_1(\gamma)=\mu$ and $\pi_2(\gamma)=\nu$
where $\pi_1$ is the projection on the first component and
$\pi_2$ is the projection on the second component.

If the probability measure~$\mu$ can be written as $d\mu=f(x)\,dx$ (i.e. it is absolutely continuous with respect to the
Lebesgue measure), then the optimal values of both problems coincide~$M_p(\mu,\nu)=W_F(\mu,\nu)$,
and there exists an optimal transport map from~$\mu$ to $\nu$
which is unique $f$-a.e. if $p>1$.
Another important characteristic of this relaxation is that it admits a dual formulation:
\begin{equation}\label{eq:dual}
W_p^p(\mu,\nu)=\sup\left\{\int_\Omega u\,d\mu+\int_\Omega v\,d\nu: 
\begin{array}{c}
u(x)+v(y)\le\lvert x-y\rvert^p\quad\textrm{for}\,\mu\textrm{-a.e.}\,x\,\textrm{and}\,\nu\textrm{-a.e.}\,y\\
u\in L^1_\mu(\Omega),v\in L^1_\nu(\Omega)
\end{array}
\right\}.
\end{equation}
Moreover, there exists an optimal pair $(u,v)$ for this dual formulation, and
when $\nu$ is an atomic probability measure (it can be written as $\nu=\sum_{i\in\mathbb{N}}b_i\delta_{y_i}$) the dual formulation becomes
\begin{equation}
W_p^p(\mu,\nu)=\sup\left\{\int_\Omega u\,d\mu+\sum_{i\in\mathbb{N}} b_i v(y_i):
\begin{array}{c}
u(x)+v(y_i)\le\lvert x-y_i\rvert^p\quad\textrm{for}\,\mu\textrm{-a.e.}\,x\,\textrm{and every}\,i\in\mathbb{N}\\
u\in L^1_\mu(\Omega),v\in L^1_\nu(\Omega)
\end{array}
\right\}.
\end{equation}
There exists another interesting characteristic when one of the measures is absolutely continuous with
respect to the Lebesgue measure and the other measure is an atomic measure.
If the probability measure~$\mu$ can be written as $d\mu=f(x)\,dx$ where~$f$ is a nonnegative function,
$(y_i)_{i\in\mathbb{N}}$ is a sequence of points in the domain~$\Omega$ such that $\nu=\sum_{i\in\mathbb{N}} b_i\delta_{y_i}$,
$(B_i)_{i\in\mathbb{N}}$ is a partition of the domain such that
the map~$T(x)=\sum_{i\in\mathbb{N}} y_i\mathbf{1}_{B_i}(x)$ is an optimal transport map from~$\mu$ to~$\nu$,
the pair $(u,v)$ is a solution of the dual formulation~\eqref{eq:dual}, then
\begin{equation}\label{eq:optim}
u(x)=\inf_{i\in\mathbb{N}} (\lvert x-y_i\rvert^p-v(y_i))=\sum_{i\in\mathbb{N}} (\lvert x-y_i\rvert^p-v(y_i))\mathbf{1}_{B_i}(x)\,\textrm{for}\,f\textrm{-a.e.}\,x.
\end{equation}
We also have a similar converse characteristic.
If $(B_i)_{i\in\mathbb{N}}$ is a partition of~$\Omega$,
we set $b_i=\int_{B_i}f(x)\,dx$, $\nu=\sum_{i\in\mathbb{N}}b_i\delta_{y_i}$ and
$T(x)=\sum_{i\in\mathbb{N}}y_i\mathbf{1}_{B_i}(x)$,
and there exists two functions $u\in L^1_\mu(\Omega)$, $v\in L^1_\nu(\Omega)$ satisfying
the condition~\eqref{eq:optim}, then $T$ is optimal for~$M_p(\mu,\nu)$
and the pair~$(u,v)$ is optimal for the dual formulation~\eqref{eq:dual}.

In order to prove the existence and uniqueness of the solution we need to consider the following:
We denote by~$S$ the unit simplex in~$\mathbb{R}^k$:
\begin{equation}
S=\left\{c=(c_1,\ldots,c_k)\in\mathbb{R}^k:\, c_i\geq 0,\,\sum_{i=1}^k c_i=1\right\}.
\end{equation}

From Theorem~\ref{theo:existenceuniqueness}, we deduce that
{\small
\begin{gather}
\inf_{\substack{ (A_i)_{i\in\{1,\ldots,k\}}\\ \textrm{partition of}\,\Omega }}
\left\{
\sum_{i=1}^k\int_{A_i}\left[
\lvert x-x_i \rvert^p+s_i\left(\int_{A_i}f(x)\,dx\right)
\right]f(x)\,dx
\right\}
=\inf_{c\in S}\left\{W_p^p(f,\sum_{i=1}^k c_i\delta_{x_i})+\sum_{i=1}^k s_i(c_i)c_i\right\}.
\end{gather}
}
% But this holds by Theorem 2.1
% and by Remark 2.2 of~\cite{crippa}.

% In Proposition~3.5 by replacing $\lvert x-x_j\rvert$ by $F(d_j(x,y))$
% we obtain the thesis.

The function $F:S\to\mathbb{R}$ defined by $F(c_1,\ldots,c_k)=W_p^p\left(\mu,\sum_{i=1}^k c_i\delta_{x_i}\right)$ is continuous and convex.

If the functions~$s_i$ are lower semi-continuous, then there exists an optimum
and if in addition the maps~$\eta_i(t):=t h_i(t)$ are strictly convex, the optimum is unique.

We can then characterize the solution. If~$(A_i)_{i=1,\ldots,k}$ is an optimum,
and $s_i$ are differentiable in~$]0,1]$ and continuous in~$0$, then the following holds:
\begin{align}
C_i =&\Big\{x\in\Omega\ : \lvert x-x_i\rvert^p+s_i(N_i)+N_i s_i'(N_i)\nonumber\\
 &\leq \lvert x-x_j\rvert^p+s_j(N_j)+N_j s_j'(N_j),\, \forall j\neq i\Big\}\\
N_i =&\int_{C_i} \lambda(x)\,dx.
\end{align}

%  {\bf Proof.-}
%  \[
%  \inf_{c\in S}\left\{W_p^p\left(f,\sum_{i=1}^k c_i\delta_{x_i}\right)\times\sum_{i=1}^k h_i(c_i)c_i\right\}
%  \]
%
%  \[
%  =\inf_{c\in S}\Big\{F(c_1,\ldots,c_k)\times\sum_{i=1}^k\eta_i(c_i)\Big\}.
%  \]
%
%  The expression to be minimized of the above equality is a lower semi-continuous function
%  defined on the non-empty compact set~$S$, hence there exists a minimizer~$(c^*_i)_{i=1,\ldots,k}$.
%
%
%
%  By Theorem 2.1 there exists an optimal transport map $T^*$ from $f$
%  to $\sum_{i=1}^k c^*_i\delta_{x_i}$.
%
%  Thanks to Remark 2.2. the transport map~$T^*$ is associated to a partition~$A^*_i$
%  of~$\Omega$ which satisfies $c^*_i=\int_{A^*_i}$ f(x)\,dx.
%
%  Then we check that this partition is an optimum.

\section*{Appendix B}
%\begin{theorem}\label{theo:multiplication}
Consider the problem~$\mathrm{(P2)}$
\begin{gather}
\operatorname*{Min}_{C_i} \sum_{i=1}^K \int\!\!\!\!\int_{C_i} \left[F(d_i(x,y))\cdot m_i\left(\int\!\!\!\!\int_{C_i}
\lambda(\omega,z)\,d\omega\,dz
\right)\right]
\cdot\lambda(x,y)\,dx\,dy,
\end{gather}
\noindent where~$C_i$ is the cell partition of~$\Omega$. Suppose that~$m_i$ are derivable.
The problem~$\mathrm{(P2)}$ admits a solution that verifies
\begin{equation}
\mathrm{(S2)}\left\{
\begin{array}{ll}
C_i=&\{x\,:\,m_i(N_i) F(d_i(x,y))\,\lambda(x,y)+U_i(x,y)\\
  &\leq m_j(N_j) F(d_j(x,y))\,\lambda(x,y)+U_j(x,y)\}\\
U_i=&m'_i(N_i)\int\!\!\!\int_{C_i}F(d_i(x,y))\lambda(x,y)\,dx\,dy\\
N_i=&\int\!\!\!\int_{C_i} \lambda(\omega,z)\,d\omega\,dz.
\end{array}
\right.
\end{equation}
%\end{theorem}

{\bf Proof.-} The proof is similar to Appendix A.
The problem~$\mathrm{(P2)}$ can be rewriten as follows:
\begin{equation}
\inf\left\{W_{c_i}\left(f\,dx,\sum_{i=1}^k c_i\delta_{x_i}\right):\, c_i\ge0,\,\sum_{i=1}^k c_i=1\right\},
\end{equation}
where
\begin{equation}
W_{c_i}\left(f\,dx,\sum_{i=1}^k c_i\delta_{x_i}\right)=\inf_{(A_i)_{i=1}^k}\left\{\sum_{i=1}^k\int_{A_i} F(\lvert x-x_i\rvert)\cdot m_i(c_i)f(x)\,dx:
\int_{A_i}f\,dx=c_i,(A_i)_{i=1}^k\,\textrm{partition of}\,\Omega\right\}.
\end{equation}
Let $\gamma$ be optimal for Monge-Kantorovich problem, then one can consider
\begin{equation}
\gamma(x,y)=\sum_{i=1}^k\gamma_i(x)\otimes\delta_{x_i}(y),\quad c_i=\gamma_i(\Omega),
\end{equation}
where $\gamma_i$ is a positive measure such that $\gamma_i\le f\,dx$

The function $F:S\to\mathbb{R}$ defined by $F(c_1,\ldots,c_k)=W_p^p\left(\mu,\sum_{i=1}^k c_i\delta_{x_i}\right)$ is lower semi-continuous.
Then there exists a solution and it is unique almost surely.

Let~$(c_i)_{i=1}^k$ be a solution of~$\mathrm{(P2)}$ and $(A_i)_{i=1}^k$ the associated optimal partition.
Let us fix two indices~$i_0$, $j_0$ and a point $x_0\in A_{i_0}$. Let $\varepsilon\ge0$.
Let us consider the open ball of radius $\varepsilon$ and center~$x_0$ that we denote as~$B_\epsilon$. We denote its measure as~$c_\epsilon$.
We make a small variation of the optimal partition by taking from~$A_{i_0}$ the ball~$B_\epsilon$ and adding it to~$A_{j_0}$.
Since the partition~$(A_i)_{i=1}^k$ is optimal then
\begin{gather}
\int_{A_{i_0}}F(\lvert x-x_{i_0}\rvert)h_{i_0}(c_{i_0})f(x)\,dx+\int_{A_{j_0}}F(\lvert x-x_{j_0}\rvert)h_{j_0}(c_{j_0})f(x)\,dx\nonumber\\
\le\int_{A_{i_0}\setminus B_\epsilon}F(\lvert x-x_{i_0}\rvert)h_{i_0}(c_{i_0}-c_\epsilon)f(x)\,dx+\int_{A_{j_0}\cup B_\varepsilon}F(\lvert x-x_{j_0}\rvert)h_{j_0}(c_{j_0}+c_\varepsilon)f(x)\,dx,
\end{gather}
which is equivalent to
\begin{gather}
\int_{A_{i_0}}F(\lvert x-x_{i_0}\rvert)(h_{i_0}(c_{i_0})-h_{i_0}(c_{i_0}-c_\epsilon))f(x)\,dx+\int_{B_\epsilon}F(\lvert x-x_{i_0}\rvert)h_{i_0}(c_{i_0}-c_\epsilon)f(x)\,dx\nonumber\\
\le\int_{A_{j_0}}F(\lvert x-x_{i_0}\rvert)(h_{j_0}(c_{j_0}+c_\epsilon)-h_{j_0}(c_{j_0}))f(x)\,dx+\int_{B_\varepsilon}F(\lvert x-x_{j_0}\rvert)h_{j_0}(c_{j_0}+c_\varepsilon)f(x)\,dx.
\end{gather}

Dividing the previous equation by~$c_\epsilon$ and taking the limit when~$\epsilon\to0$, we obtain
\begin{gather}
\int_{A_{i_0}}F(\lvert x-x_{i_0}\rvert)h'_{i_0}(c_{i_0})f(x)\,dx+F(\lvert x-x_{i_0}\rvert)h_{i_0}(c_{i_0})f(x_0)\nonumber\\
\le\int_{A_{j_0}}F(\lvert x-x_{j_0}\rvert)h'_{j_0}(c_{j_0})f(x)\,dx+F(\lvert x-x_{j_0}\rvert)h_{j_0}(c_{j_0})f(x_0).
\end{gather}

Reorganizing the terms we obtain the desired result.

\bibliographystyle{hieeetr}
%\bibliography{IEEEabrev,mybibfile}
\bibliography{mybibfile}

\end{document}